\documentclass[a4paper,11pt]{article}
\pdfoutput=1 % if your are submitting a pdflatex (i.e. if you have
\pdfoutput=1 % if your are submitting a pdflatex (i.e. if you have
\usepackage{jheppub} % for details on the use of the package, please
\usepackage[T1]{fontenc} % if needed
\def\0{{\bf 0}}
\def\e{{\bf e}}
\def\r{{\bf r}}
\def\m{{\bf m}}
\def\l{\ell}
\def\mm{{\rm m}}

\title{\boldmath  Entanglement entropy of gravitational edge modes}

 \title{\textbf{\textsf{ Entanglement entropy of gravitational edge modes
}}}
  \author{Justin R. David, Jyotirmoy Mukherjee}
\affiliation{\vspace{.1cm} Centre for High Energy Physics, \\ Indian Institute of Science,\\
C. V. Raman Avenue, Bangalore 560012, India.}
\emailAdd{justin@iisc.ac.in, jyotirmoym@iisc.ac.in}

\abstract{
We  consider the linearised graviton in $4d$ Minkowski space and decompose it into tensor spherical harmonics and fix the gauge. The Gauss law of gravity implies that certain radial components of the Riemann tensor 
of the graviton on the sphere labels the superselection sectors for the graviton.  
We show that 
among these  6 normal  components of the Riemann tensor, 2 are related locally to the 
algebra of gauge-invariant operators in the sphere.
From the two-point function of these  components of the Riemann tensor on $S^2$  we compute the 
logarithmic coefficient  of   the entanglement entropy of these superselection sectors  across 
a spherical entangling surface.  
For sectors labelled by each of the two components of the Riemann 
tensor these coefficients  are equal and their total contribution is given by $-\frac{16}{3}$.  
We observe that this coefficient
coincides with
 that extracted from the edge partition function of the massless spin-2 field on the 4-sphere when written in terms of its Harish-Chandra character.
As a preliminary step, we also evaluate the logarithmic coefficient of 
 the entanglement entropy from 
the superselection sectors labelled by the radial component of the electric field
of the $U(1)$ theory in even $d$ dimensions. 
We show that this agrees with the corresponding coefficient of 
the edge Harish-Chandra character 
of the massless spin-1 field on $S^d$. 
}

\begin{document} 
\maketitle
\flushbottom

\section{Introduction}

Entanglement entropy  in theories with local gauge symmetries is difficult to define. In order to define entanglement entropy, the surface at a constant time slice is sub-divided into a region 
$A$ and its complement $\bar A$. We then require that the Hilbert space of the theory naturally 
factorizes as
\begin{equation}
{\cal H} = {\cal H}_A \otimes {\cal H}_{\bar A}.
\end{equation}
Scalar and spinor field theories have local physical excitations and their Hilbert space admits such factorization. 
In this factorized Hilbert space, one defines the reduced density matrix by tracing over 
${\cal H}_{\bar A} $  which we denote by  $ \rho_A = {\rm Tr}_{\bar A} \rho $. 
Then the entanglement entropy of the region $A$ is given  by 
\begin{equation}
S(\rho_A ) = -{\rm Tr}_{ A} (\rho_A \log \rho_A).
\end{equation}

Theories with gauge symmetries including quantum gravity do not always
admit local gauge-invariant operators and therefore the gauge-invariant Hilbert space does not admit a natural tensor product structure. For gauge theories, this issue was discussed in 
\cite{Buividovich:2008gq,Donnelly:2011hn,Casini:2013rba,Radicevic:2014kqa,Casini:2014aia,Donnelly:2014gva,Huang:2014pfa,Donnelly:2014fua,Ghosh:2015iwa,Aoki:2015bsa,Donnelly:2015hxa,Soni:2015yga,VanAcoleyen:2015ccp,Casini:2015dsg,Radicevic:2015sza,Soni:2016ogt} and has been summarized recently  in \cite{Casini:2022rlv}. 
This lack of factorization of the Hilbert space is due to the presence of a centre of the algebra of operators. 
The operators belonging to the centre commute with all others and a unique  entropy can be assigned 
to the centre.  This centre can be chosen  in many ways \cite{Casini:2013rba}. A choice which arises naturally 
when  one considers the extended Hilbert space description of lattice gauge theories  is 
called the electric centre \cite{Ghosh:2015iwa}.
This non-trivial centre  occurs due to the electric Gauss law constraint on physical states.
Similarly, the magnetic Gauss law constraint gives rise to the magnetic centre. 
Such choices of centres also exist for scalar field theories 
on a lattice in which keeping the field itself fixed on the boundary leads to a non-trivial centre and a contribution 
to entanglement entropy.
It is also possible to choose the algebra  on lattice gauge theories such that the centre is trivial 
and therefore there is no entropy associated with it \cite{Casini:2013rba}. 
In this paper, we will evaluate the entanglement entropy associated with centres that arise 
naturally from the Gauss law constraints for linearised gravity. 
At this point, the reader might wonder, why is it that we are focused on a quantity that is 
non-universal and depends on the choice of the centre. 
As we will see, the  logarithmic coefficient corresponding to the 
 centre obtained naturally from the Gauss law constraints
in linearised gravity coincides with
 that extracted from the edge partition function of the massless spin-2 field on the 4-sphere when written in terms of its Harish-Chandra character. Furthermore, we feel that  the techniques developed in this paper are useful and applicable 
 to other theories with local symmetries.

To begin, let us consider the $U(1)$ theory, 
since physical states obey the Gauss law constraint and gauge-invariant operators do not change the 
electric flux normal to the entangling surface, the Hilbert space ${\cal H}_A$ as well as ${\cal H }_{\bar A}$ 
 factorizes into superselection sectors labelled by the electric flux normal to the entangling surface.
  Then the entanglement entropy for a theory with  $U(1)$  gauge symmetry is given by
  \begin{equation} \label{eesplit}
S(\rho_A)=-\sum_E  p_E \log(p_E)+ \sum_E p_E S(\rho_A^E ) .
\end{equation}
Here $p_E$ is the probability associated with a given superselection sector labelled by the electric flux. 
The first term in this expression is just the Shannon or classical entropy associated with the superselection sector. This contribution is also referred to as the entanglement entropy of edge modes and its contribution is non-extractable. That is, this entropy cannot be distilled into a number of Bell pairs \cite{Soni:2016ogt}.  This term is the entanglement contribution 
of the electric centre. 
 In \cite{Donnelly:2014fua,Huang:2014pfa,Soni:2016ogt}, 
  the entanglement entropy of electromagnetic edge modes was evaluated for 
  a spherical entangling surface of a $U(1)$
   theory in 4-dimensions and it was shown that it is captured by the partition function of a massless scalar on $S^2$.

 As mentioned earlier,   
the main aim of this paper is to evaluate the contribution of the entanglement entropy of the edge modes of the linearised graviton. This theory can be treated as a quantum field theory of spin-2 particles and therefore the question of whether local subsystems exist in the full quantum theory of gravity does not arise. 
In a certain gauge the linearised graviton $h_{\mu \nu}$, 
 can be algebraically related to the curvature which is gauge-invariant and it generates the algebra of gauge-invariant operators of this theory. In  \cite{Benedetti:2019uej}, 
  this approach was used to evaluate the logarithmic term in the entanglement entropy of linearised graviton across a spherical entangling surface. The Gauss law of the theory implies that the Hilbert space decomposes 
  as a sum of superselection sectors similar to 
  (\ref{eesplit}). 
  We will show that the superselection sectors in this case are labelled by the normal components of the Riemann tensor on the sphere. Using this we will evaluate the classical non-extractable 
contribution of the entanglement entropy of the edge modes of the graviton.
  
A particularly direct method of evaluating the entropy of the edge modes for the Maxwell theory involves constructing the probability distribution $p_E$ of the superselection sectors using the two-point function of the normal component of the electric field on the sphere. This was developed in \cite{Soni:2016ogt}. 
We revisit this computation  before we proceed to the graviton. In \cite{Soni:2016ogt}, 
the Maxwell theory was quantised in cartesian coordinates. We find it convenient to expand the $U(1)$
 field in vector spherical harmonics, fix gauge and quantise the theory following \cite{Casini:2015dsg,Benedetti:2019uej}. Then we show that the radial component of the electric field on the sphere is directly related to one of the two canonical momenta. This allows us to evaluate the two-point function of the electric field on the sphere  and compute the contribution of the edge modes. We show that indeed the logarithmic coefficient of the entanglement entropy of the edge modes 
 or the electric centre is 
 obtained from the corresponding coefficient of 
 the partition function of the massless scalar on 
 the sphere $S^2$. To demonstrate the utility of this approach, we generalise the computation to the 
 $U(1)$
  theory in arbitrary  even $d$-dimensions. We show that the logarithmic coefficient of the 
  non-extractable contribution to the entanglement entropy can be obtained from the 
  corresponding coefficient of 
  the partition function of a massless scalar on $S^{d-2}$. 
Recently it has been shown that the partition function of the vector on $S^d$ when written as an integral over the Harish-Chandra character naturally decomposes into a sum of contributions from bulk and edge characters \cite{Anninos:2020hfj}. 
We observe that the contribution from the edge characters precisely coincides with that non-extractable entanglement entropy corresponding to the superselection sectors determined by the Gauss law. 
  
We then consider the linearised graviton, we first demonstrate that the Gauss law of gravity implies that the 
certain
normal
components of the Riemann tensor to the entangling surface labels the superselection sectors. 
These components have one time direction, one radial direction and the rest arbitrary, there are $6$ such components.
Then we briefly review the quantization of the graviton developed in \cite{Benedetti:2019uej}. Here the field $h_{\mu\nu}$
 is expanded in terms of tensor harmonics and under an appropriate gauge choice leads to a pair of canonical coordinates and momenta. We show that among the $6$ components of the Riemann tensor, 
 only  two 
are  algebraically or locally related to the canonical coordinates.
We choose these  components of the curvature tensor to label the superselection sectors and
evaluate  their two-point function on $S^2$. From this we evaluate the 
contribution of the superselection sectors to the non-extractable entanglement entropy.  We see that
the logarithmic coefficient of the classical entanglement of the superselection sectors labelled
by each of the 2 Riemann tensors are equal and given by $-8/3$. 
Since the 2 Riemann tensors are independent labels of the superselection sectors, their sum represents
the complete contribution of the edge modes due to the centre obtained due to the  Gauss law constraints.
Then we observe that just as in 
the $U(1)$ case, we see that the logarithmic term of this contribution of  both the superselection sectors
of the 
graviton  $-16/3$, 
 agrees with the  term  from edge mode partition function of the massless spin-2 field on $S^4$. Here the edge mode partition function is identified by writing the 
spin-2 partition function on $S^4$ as an integral over its Harish-Chandra character.

The organisation of the paper is as follows.
In section \ref{u1edge} we 
re-visit the evaluation of the classical entropy of the superselection sectors for the $U(1)$ theory. 
This is first done in $4d$ and then in arbitrary even dimensions. 
In section \ref{gravsection}, we evaluate the contribution of the superselections sectors or in other words the 
electric centre 
 to the  logarithmic coefficient 
of the entanglement entropy of the linearised graviton across a spherical surface.  
Section \ref{curvcomp} contains the evaluation of the 6 components of the curvature which 
must obey the Gauss law. The explicit computation reveals 
only 2 of these are locally related to the algebra of gauge-invariant operators in a sphere. 
Section \ref{conclusion} contains our conclusions. 
The appendix  \ref{appen}, 
compares the two-point function of the radial components of the 
electric field on $S^2$ evaluated in the Coulomb gauge evaluated in \cite{Soni:2016ogt} with that evaluated 
using the expansion in vector harmonics and the gauge introduced in \cite{Casini:2015dsg,Benedetti:2019uej}. 
This paper uses the latter gauge which is more suited to the spherical symmetry of the problem 
and which can be generalised for the graviton.

\section{$U(1)$ edge modes} \label{u1edge}

In this section we begin with the review  of the approach of \cite{Soni:2016ogt} to evaluate the
contribution of the  entanglement entropy of the superselection sectors of the 
$U(1)$ gauge field in even $d$ dimensions. 
Since we work with a spherical entangling surface it is convenient to 
use the methods of \cite{Casini:2015dsg,Benedetti:2019uej} to quantize the system. 

\subsection{Entanglement from correlators  on the sphere}
\label{sec2.1}

Consider the $U(1)$ theory with the action  given by 
\begin{eqnarray} \label{u1act}
S = - \frac{1}{4} \int d^d x   F_{\mu \nu} F^{\mu\nu}, 
\end{eqnarray}
and the spherical entangling surface $S^{d-2}$.   We restrict our attention to even $d$ since 
our focus is to obtain the  logarithmic coefficient of the entanglement entropy. 
All physical states satisfy the Gauss law constraint 
\begin{eqnarray}\label{u1glaw}
\partial^\mu F_{\mu 0 } =0.
\end{eqnarray}
This implies that the normal  component of the electric field $F_{0r}$ has to  match 
across the entangling surface. Gauge-invariant operators like Wilson lines acting inside 
the entangling region $A$ or  those outside in $\bar A$ cannot change this electric field. 
This leads to the factorisation of the density matrix into superselection 
sectors labelled by the normal component of the electric field. 
The classical contribution to the entanglement entropy from these superselection 
sectors is given by 
\begin{equation}
S_{{\rm edge} } ( \rho_A) = -\sum_{E }p(E) \log  p (E) . 
\end{equation}
where $p_E$ is the probability associated with a given superselection sector.

The $U(1)$ theory is free and therefore  $p(E)$ is  a  Gaussian functional of 
the normal component of the electric field given by
\begin{equation}\label{prob1}
p[E_r] = {\cal N} \exp\left[ 
 -\frac{1}{2} \int d^{\hat d} xd^{\hat d} x^\prime E_{r}( x)  G_{rr'}^{-1} ( x, x')  E_{r' } (x') \right].
\end{equation}
Here $x, x'$  are coordinates on the $\hat d \equiv d-2$ sphere,  $S^{\hat d}$. 
$G_{rr'} (x , x')$ is the  two-point correlator 
of the radial component of the electric field on the sphere which is defined by 
\begin{equation} \label{elec2pt}
G_{rr'} (x, x') = \langle 0 | F_{0 r'} (x) F_{0 r'} ( x' ) |0 \rangle, 
\end{equation}
and its inverse satisfies the equation
\begin{equation} \label{prob2}
\int d^{\hat d} x'' G_{rr''} ( x,  x'') G^{-1}_{r''r'} ( x'' , x') = \delta^{\hat d } ( x-x').
\end{equation}
Note that  the two-point function in (\ref{elec2pt}) are evaluated on the entangling sphere, 
the labels $r, r'$, just refers to the fact  the correlator is between the radial components 
at angular locations $x, x'$, on the sphere. 
The integrals in (\ref{prob1}) and (\ref{prob2}) are over  the sphere $S^{\hat d}$

Evaluating the classical contribution 
we get 
\begin{eqnarray}
S_{{\rm edge} } ( \rho_A)  = - \log {\cal N }+ \int d^{\hat d} x d^{\hat d} x' 
G_{rr'} ( x, x') G^{-1}_{r'r}( x', x) .
\end{eqnarray}
Using (\ref{prob2}), we see that the term involving the integrals is divergent
\begin{equation}
\int d^{\hat d} x d^{\hat d } x' 
G_{rr'} ( x, x') G^{-1}_{r'r}( x', x)  = \int d^{\hat d} x \delta^{\hat d} (0) .
\end{equation}
We can regulate the delta function by introducing a cut-off $\epsilon$, this cut off 
is a short distance cut off along the angular directions on   surface of the sphere. 
Therefore we can write
\begin{equation}
\int d^{\hat d} x d^{\hat d} x' 
G_{rr'} ( x, x') G^{-1}_{r'r}( x', x)  = \frac{ R^{\hat d } {\rm Vol} ( S^{\hat d})  }{ \epsilon^{\hat d}} .
\end{equation}
Here $R$ is the radius of the entangling surface and ${\rm Vol} (S^{\hat d}) $ is the volume of 
the unit $S^{\hat d}$ sphere. 
This term is proportional to the area and does not contribute to the  
 logarithmic term which  is proportional to $\log\frac{ R}{\epsilon}$.
Let us now study the term ${\cal N}$. 
From the condition  
\begin{equation}
\int {\cal D} E_r \; p[E_r]  = 1,
\end{equation}
we obtain 
\begin{equation}
\log {\cal N} -  \frac{1}{2} \log\big( {\rm det }\,  G^{-1}_{r r'} \big)  =0.
\end{equation}
We have 
 defined the measure  of the functional integral by the following integral
\begin{equation}
\int{\cal D} E_r \exp\left[  - \frac{1}{2} \int d^{\hat d} x E_r^2 (x)  \right]  =1.
\end{equation}
Therefore the   contribution to the entanglement of the superselection sectors
is given by 
\begin{equation} \label{unived}
\left. S_{{\rm edge} } ( \rho_A) = \frac{1}{2} \log {\rm det} G_{rr'}\right|_{\log \; {\rm coefficient}}.
\end{equation}
Here we have implicitly assumed that is it only the logarithmic contribution to 
$S_{\rm edge}$ we are interested in. 

The two-point function of the electric field with both the electric fields 
on the same sphere diverges, as we will see subsequently. 
To regulate this divergence we consider the correlator where the radial 
component of the electric fields lie on two spheres of radius $r$ and $ r'= r + \delta$
\begin{equation}\label{splitgrr}
G_{r r' } ( r, r'; x,  x')  = \langle 0|  F_{0 r} ( r, x) F_{0r'} ( r'. x') |0 \rangle.
\end{equation}
We have introduced $r, r'$ in the arguments of the Greens function to make it explicit that 
the  electric field correlator involves insertion on spheres of different radii. 
We need to take $\delta \rightarrow 0$ such that 
\begin{equation}
\delta << \epsilon,
\end{equation}
where $\epsilon$ is the short distance cut-off on the sphere. 

Therefore we look for the leading divergence  in (\ref{splitgrr}) 
when $\delta\rightarrow 0$ and use its coefficient in (\ref{unived}) 
to obtain the  logarithmic term proportional to $\log\frac{r}{\epsilon}$. 
In the section (\ref{u1arbd})  we  will evaluate the two-point function of the radial component 
of the electric field in the angular momentum basis on $S^{\hat d}$. 
We  show that  in the $\delta\rightarrow 0$ limit, the two-point function admits the expansion
\begin{equation} \label{deltalim}
\lim_{\delta \rightarrow 0} \langle \l \lambda | G_{r r' } ( r, r + \delta  x,  x') | \l' \lambda'  \rangle   
 = \frac{ \l( \l +d -3) }{ 4 \pi r^d}  \Big( \log \frac{ r^2}{ \delta^2 }  \Big)  \delta_{\l, \l' } \delta_{\lambda, \lambda'}   + O(\delta^0) .
 \end{equation}
Here $\l$ labels the eigen value of the Laplacian of scalars, and $\lambda$ refers to all the 
other quantum numbers of the scalar harmonics on $S^{\hat d}$. 
Note that $\l ( \l +d-3)$ is the eigenvalue of the scalar Laplacian on $S^{\hat d}$. 
Then substituting the coefficient of the leading divergence in (\ref{deltalim}) we see that 
the  coefficient to the 
entanglement entropy of the superselection sectors is given by 
coefficient of the logarithmic divergence of the one-loop determinant of the massless scalar
on $S^{\hat d}$. Since the cut off on the surface of the sphere is $\epsilon$, this divergence is 
proportional to $\log \frac{ R}{\epsilon}$.

\subsection{The Maxwell theory in $d=4$}

In this  section, we first briefly discuss the method introduced by \cite{Benedetti:2019uej} 
to quantise the photon in 
spherical coordinates. 
We introduce the covariant notation for the vector harmonics
 which enables evaluation of field strengths easily, this notation will be carried over to the discussion 
 of the graviton. 
 After fixing gauge and setting up the canonical commutation relations for the gauge-invariant 
 conjugate variables, we evaluate the two-point function of the radial component of the 
 electric field
 
 We first expand the vector potential $A_\mu$ as follows
 \begin{eqnarray} \label{defexpa}
 A_\mu = \sum_{\l, m } \left(  A_{{ 0} (\l , m)} ( r, t) Y_{l m, \mu  }^{ \0} + 
 A_{{ r}  (\l,  m)  } ( r, t) Y^{ \r}_{\l m, \mu }
 + 
 A_{{ e} ( \l,  m)  } ( r, t) Y^{ \e} _{\l m , \mu }
 +
 A_{{ \mm} (  \l,  m)  } ( r, t) Y^{ \m}_{\l m ,\mu} \right),  \nonumber
 \\
 \end{eqnarray}
 here the greek subscript  refers to the component of 
  the covariant vectors which are defined as follows
 \begin{eqnarray}\label{defcov}
 Y_{\l m }^{\0}  &=& \left\{ Y_{\l m} ( \theta, \phi) , 0 , 0, 0 \right\}, \\ \nonumber
Y^{\r}_{\l m} &=& \left\{ 0, Y_{\l m} ( \theta, \phi), 0, 0 \right\} , \\ \nonumber
 Y^{\e} _{\l m }
&=&\frac{r }{\sqrt{ \ell  (\ell +1)}}
\left\{0,0,\frac{\partial Y_{\l m}(\theta ,\phi )}{\partial \theta },\frac{\partial Y_{\l m} (\theta ,\phi )}{\partial \phi }\right\},
\\ \nonumber
              Y^{\m}_{\l m} &=&\frac{r}{\sqrt{\ell  (\ell +1)}}
               \left\{0,0,-  \frac{1}{\sin\theta}  \frac{\partial Y_{\l m }(\theta ,\phi )}{\partial \phi },\sin (\theta ) \frac{\partial Y_{\l m} (\theta ,\phi )}{\partial \theta }\right\}.
 \end{eqnarray}
 $Y_{\l m} ( \theta, \phi)$ are scalar spherical harmonics with $\l =0, 1, 2 \cdots$ and $-\l\leq m \leq \l $. 
 In (\ref{defcov}) wherever derivatives of the spherical harmonics occur it is understood that 
  $\l = 1, 2, \cdots$.  
 We have converted the conventional vector harmonics used in \cite{Benedetti:2019uej} which are vectors whose 
 inner product is defined by the dot product using Kr\"{o}necker delta to covariant vectors. 
 This makes it easy to apply the methods of covariant tensor calculus rather than 
 vector calculus.  The metric is the flat space metric written in polar coordinates
 \begin{equation} \label{flat}
 ds^2 = - dt^2 + dr^2 + r^2 ( d\theta^2 + \sin^2 \theta d\phi^2) .
 \end{equation}
 Raising and lowering indices as well as covariant derivatives are defined with 
 respect to this metric. 
 It is also useful to introduce covariant  orthonormal vectors in  the directions of vectors in (\ref{defcov}). 
 \begin{eqnarray}\label{unitvec}
 \hat t  &=&    \{ 1, 0, 0, 0 \},  \\ \nonumber
 \hat r &=& \{ 0, 1, 0, 0 \}, \\ \nonumber
 \hat e &=& \frac{ Y^\e_{lm}}{ | Y^\e_{\l m}  |}, \qquad\qquad |Y^\e _{\l m} |^2 =  g^{\mu\nu} Y^\e_{\l m;\mu} 
 Y^\e_{\l m;\nu},  \\ \nonumber 
 \hat m &=&  \frac{ Y^\m_{\l m} }{ | Y^\m_{\l m} |}, \qquad\qquad |Y^\m_{\l m} |^2 = 
  g^{\mu\nu} Y^\m_{\l m; \mu }Y^\m_{\l m ;\nu}, 
 \end{eqnarray}
 where the metric $g_{\mu\nu}$ is given in (\ref{flat}). Projections of tensors along these 
 directions are defined as follows, consider a covariant tensor $T_{\mu\nu\rho}$, then 
 \begin{equation}
 T_{\hat e \hat m \hat r } = \hat e^\mu\hat m^\nu \hat r^\rho T _{\mu\nu\rho}.
 \end{equation}
 Similar definitions apply for other projections.   The reality of the vector potential implies the following 
 reality property of the coefficients of the expansion in (\ref{defexpa}). 
 \begin{eqnarray}\label{reality}
 && (-1)^m A_{0, (l, m ) } ^*  = A_{0, (l, -m ) }, \qquad
 (-1)^m A_{r, (l, m ) } ^*  = A_{r, (l, -m ) }, \\ \nonumber
  && (-1)^m A_{e, (l, m ) } ^*  = A_{e, (l,  -m ) }, \qquad
   (-1)^m A_{\mm, (l, m ) } ^*  = A_{\mm, (l,  -m ) }.
 \end{eqnarray}
 Let us also expand the gauge transformation in terms of spherical Harmonics by 
 \begin{equation}
 \chi = \sum_{\l, m} \chi_{\l m } (r, t) Y_{\l m} ( \theta, \phi) .
 \end{equation}
 Then the gauge field transforms as 
 \begin{eqnarray}
  A_\mu &=& \sum_{\l, m } \Big(  \big[ A_{{ 0} ( \l,  m) } ( r, t)  + \dot \chi_{\l m } ( r, t) \big]Y_{\l m; \mu  }^{\bf 0} + 
 \big[ A_{{ r} (  \l,  m)  } ( r, t)  + \partial_r  \chi_{\l m}( r, t)  \big]Y^{\bf r}_{\l m; \mu }   \\ \nonumber 
& &  \big[  A_{{ e} ( \l,  m)  } ( r, t)  + \frac{\chi_{\l m} (r, t) }{r} \big]Y^{\bf e} _{\l m ; \mu }
 +
  A_{{\mm} ( \l,  m)  } ( r, t)  Y^{\bf m}_{\l m ;\mu} ) \Big) .
 \end{eqnarray}
 Here the superscript  `$\;\dot{}\;$' refers to partial derivative with respect to time. 
 As in \cite{Benedetti:2019uej} we choose the gauge such that the longitudinal component of the gauge 
 field $A_{e; \l m }$ vanishes for every $\l, m $. 
 Evaluating the field strengths in this gauge we obtain
 \begin{eqnarray}\label{fieldstr}
F_{\hat t \hat r} &=& \sum_{\ell,m}(\dot{A}_{r; \ell m}-\partial_r A_{0; \ell m})Y_{\ell m} \\ \nonumber
F_{\hat t \hat m} &=&   \sum_{\ell,m} \dot A_{\mm; \ell m } |Y^{\m}_{\ell m}  |, \\ \nonumber
F_{\hat t \hat e} &=& -   \sum_{\ell,m} \frac{\sqrt{\ell ( \ell + 1) } }{r}  A_{0; \ell m}  |Y^\e_{\ell m } |, \\ \nonumber
F_{\hat r \hat e}&=&   \sum_{\ell,m} \frac{\sqrt{\ell ( \ell + 1) } }{r}  A_{r; \l m } |Y^\e_{\ell m } |, \\ \nonumber
F_{\hat r \hat m} &=& -  \sum_{\ell,m} \frac{ \partial_r ( r A_{\mm; \ell m }) }{r}  |Y^{\m}_{\ell m}  |, \\ \nonumber
F_{\hat e \hat m} &=&   \sum_{\ell,m}  \frac{ \sqrt{ \l ( \l + 1) }}{r}  A_{\mm; \ell m } Y_{\ell m} .
 \end{eqnarray}
 
 Expanding the gauge field in the action (\ref{u1act})  for $d=4$  and using the orthogonality properties 
 of spherical harmonics we obtain 
 \begin{eqnarray} \label{actionl}
S &=& -\frac{1}{4}
            \int r^2dr d\Omega F_{\mu\nu}F^{\mu\nu} =\sum_{\l , m= -\l }^\l \int dr L_{\l m} , \nonumber\\
        L_{\l m}    &=&\frac{1}{2} \Big[ r^2 \dot{A}_{r ( \ell,  m) } \dot{A}_{r (  \ell,  m )  }^*+r^2\dot{A}_{\mm ( \ell,  m) }
        \dot{A}_{\mm( \l,  m) }^*- \ell(\ell+1) A_{ r ( \ell,  m) } A_{ r ( \ell, m) }^* \nonumber\\
            & &-\ell(\ell+1)A_{\mm ( \ell,  m) }A_{ \mm ( \ell,  -m) }-|A_{\mm ( \l, m)} +r\partial_rA_{\mm (\l, m) }|^2
            +r^2 \partial_r A_{0 ( \ell,  m) } \partial_r A_{0( \ell m)}^* \nonumber\\
            & &
            -r^2\dot{A}_{ r, (\ell m) } \partial_r A_{0( \ell m) }^*  -r^2\dot{A}_{r ( \ell m) }^*\partial_r A_{0( \ell m) } +\ell(\ell+1)A_{ 0 (\l, m )} A_{0 ( \l,  m ) }^*  \Big].
            \end{eqnarray}
            The sum over $\l$ in  (\ref{actionl}) is understood to run from either $\l =0, 1\cdots$ or 
            $\l = 1, 2, \cdots$ depending on the whether the mode corresponds to scalar or vector harmonics.
            The canonical conjugate momenta to $A_r$ and $A_{\mm}$ for $m=0, 1\cdots$ are given by 
            \footnote{We have taken the contribution of both positive and negative values of $m$ and used the reality properties in writing the  canonical momenta. This removes the factor of $1/2$. For $m=0$, the field is real}
 \begin{equation} \label{ccm}
 \pi_{ (\l, m) }^{r \, *} 
 = \frac{\partial L_\l}{\partial \dot A_{ r ( \l , m ) }^* } = r^2 \left[  \dot A_{r ( \l, m ) }  - \partial_r A_{0 ( \l, m ) } \right] , 
 \qquad 
 \pi^{\mm\, *}_{( \l,  m ) } = \frac{\partial L_\l}{\partial \dot A_{ \mm ( \l, m  ) }^* } = r^2 \dot A_{\mm ( \l, m  ) } .
 \end{equation} 
 As a step towards quantization of the modes $A_r$ and $A_\mm$ we obtain the wave  equations 
 as well as the solutions 
satisfied by these modes and their conjugate variables. 

\subsubsection*{The mode $A_r$}

 From the action in (\ref{actionl}), we obtain 
 \begin{align}
 r^2 ( \ddot{A}_{r (\l, m ) } - \partial_r \dot A_{0 ( \l, m  ) } ) + l ( l + 1) A_{ r ( \l, m )}  =0, 
 \end{align}
 which can also be written as 
 \begin{equation}
\dot  \pi^{r\, *}_{( \l, m ) } + \l ( \l + 1) A_{r (\l , m ) } =0.
 \end{equation}
 We can eliminate $A_{0 (\l, m)}$ using the constraint
 \begin{equation} \label{cona0}
 \partial_r \pi^{r\, *}_{( \l, m ) } + \l ( \l + 1) A_{0 ( \l , m ) }  =0.
 \end{equation}
 This leads a closed equation for $A_r$
 \begin{equation}
 r^2 ( \ddot A_{r(\l, m ) } - \partial_r^2 A_{r(\l, m )} ) + \l ( \l + 1) A_{r (\l , m )} = 0 .
 \end{equation}
 We can solve this equation by expanding in Fourier modes in time and then solving the 
 radial equation. The solution  for a particular Fourier mode  labelled by $k$ 
 which is regular at the origin can be written as 
 \begin{equation} \label{class1}
 A_{r (\l, m ) }  = e^{- i k t}  a_{r (\l, m )}  (k )  \sqrt{r} J_{\l + \frac{1}{2} } ( |k| r) .
 \end{equation}
 Here $a_{\l, m }$ is the integration constant and $J_{\l +\frac{1}{2} }$ refers to the Bessel function.
 The equation $\pi^r_{( \l m ) }$ can be obtained using the definition in (\ref{ccm})
 \begin{eqnarray}
 \pi^{r\, *}_{(\l , m ) } &=& r^2 \left(  - i k   e^{- i k t}  a_{\l, m }  (k )   \sqrt{r} J_{\l + \frac{1}{2} } ( |k| r)  - \partial_r A_{0 ( \l m)} \right) , \\ \nonumber
 &=& r^2  \left(  - i k   e^{- i k t}  a_{r( \l, m ) }  (k )   \sqrt{r} J_{\l + \frac{1}{2} } ( |k| r)   + \frac{1}{\l ( \l + 1) }
 \partial_r^2  \pi^{r\, *} _{(\l , m ) } \right) .
 \end{eqnarray}
 In the second line we have used the constraint in (\ref{cona0}) to eliminate $A_0$. 
 Therefore the equation $\pi^r_{(\l, m ) }$ is an in-homogenous 
 second order equation in the radial coordinate. 
 The general solution   is given by 
 \begin{eqnarray}
 \pi^{r\, *} _{( \l, m )} &=&  c_1 r^{\l +1}  + c_2 r^{-\l}  + a_{ r( \l, m ) } (k) e^{- i k t} \left( 
 \frac{i 2^{-\ell-\frac{1}{2}} \ell (\ell+1)  k ^\ell}{\sqrt{k} \Gamma \left(\ell+\frac{3}{2}\right)} r^{\l+1} 
 -\frac{i \ell (\ell+1) \sqrt{r} }{k}  J_{\ell+\frac{1}{2}}(|k| r) \right).  \nonumber \\
 \end{eqnarray}
 Demanding that the solution be regular at the origin yields  $c_2 =0$ since $\l\geq 1$ \footnote{
 For $\l =0$, using  (\ref{cona0}) and demanding
  that the solution vanish at infinity we see  $\pi^{r} _{( \l, m )} =0$.}
  for these modes. 
 Further demanding that the solution be well defined at infinity fixes $c_1$ and yields
 \begin{equation} \label{class2}
  \pi^{r\, *}_{( \l, m )}  =  - \frac{i \l( l +1) }{k}  a_{ r( \l, m ) } (k) e^{- i k t}   \sqrt{r}  J_{\ell+\frac{1}{2}}(|k| r).
 \end{equation}
 The classical solutions in (\ref{class1}) and (\ref{class2}) allow us to write the mode expansion of the 
 fields $A_{r (\l, m ) }, \pi^r_{(\l, m ) }$  as follows
 \begin{eqnarray}\label{modexp1}
 A_{r ( \l. m) } ( r, t ) &=& \frac{1}{\sqrt{2} } 
 \int_0^\infty 
  k dk \left( a_{r (\l, m) } (k) e^{ - i k t}  + (-1)^m  a_{r( \l, -m )}^\dagger (k)  e^{i k t}  \right) 
  \sqrt{r} J_{\l + \frac{1}{2} } ( k r) , \\ \nonumber
 \pi^{r\, *}_{( \l, m ) } ( r, t) &=&   \frac{ \l ( \l +1) }{\sqrt{2} } 
 \int_0^\infty  dk \left( - i a_{r (\l, m) } (k) e^{ - i k t} +  i (-1)^m a_{r(\l,  -m)}^\dagger(k)  e^{i k t}    \right)\sqrt{r} J_{\l + \frac{1}{2} } ( k r) .
 \end{eqnarray}
 This  form for the mode expansion respects the reality condition (\ref{reality}).
 Let us also write the mode expansion 
 \begin{eqnarray}
 \pi^{r}_{( \l, m ) } ( r, t) &=&   \frac{ \l ( \l +1) }{\sqrt{2} } 
  \int_0^\infty  dk \left(  i a_{r (\l, m) }^\dagger (k) e^{  i k t} -  i (-1)^m a_{r(\l,  -m)}(k) e^{-i k t} 
    \right)\sqrt{r} J_{\l + \frac{1}{2} } ( k r).  \nonumber \\
 \end{eqnarray}
We can now promote $A_r, \pi^r$ to be operators which  implies that $a_{\l, m }, a^\dagger_{\l, m }$ 
are operators. 
The equal time 
canonical commutation relation of these conjugate 
variables is given by 
\begin{equation} \label{comap}
[A_{r (\l,  m ) }( r, t)  , \pi^r_{ (\l', m') } ( r', t)  ] = i \delta_{l, l'} \delta_{m, m'} \delta ( r - r') .
\end{equation}
Using the mode expansion in (\ref{modexp1}), it can be seen that this commutation relation implies 
the following commutation relations between the creation and annihilation operators
\begin{equation}\label{crean1}
[a_{r (\l, m )} (k) , a^{\dagger}_{r (\l', m')} (k') ] =\frac{1}{\l ( \l +1)}  \delta( k-k') \delta_{\l, \l'} \delta_{m, m'}.
\end{equation}
All other commutation relations are trivial. 
To  show (\ref{crean1})  holds, 
we substitute the expansion (\ref{modexp1}) in (\ref{comap}) and use the closure relation, see
\cite{arfken2005mathematical},  section 11.2, 
\begin{equation}\label{Besselclosure}
\int_0^\infty  k dk J_{\l + \frac{1}{2} }(k r)  J_{ \l + \frac{1}{2} }( k r')  = \frac{1}{r} \delta ( r- r') .
\end{equation}

\subsubsection*{The mode $A_\mm$}

From the action (\ref{actionl}), the equations of motion for the mode $A_\mm$ is given by 
\begin{eqnarray}
r^2 (\partial_r^2  A_{\mm (\l, m ) }  - \ddot{A}_{\mm (\l, m ) } )  + 2 r \partial_r  A_{\mm (\l, m ) }  - 
\l( \l +1)   A_{\mm (\l, m ) } =0.
\end{eqnarray}
The solution for each Fourier mode in time which is regular at the origin is given by 
\begin{equation}\label{amsol1}
A_{\mm (\l, m ) } = a_{\mm (\l, m) } (k) r^{-\frac{1}{2}}  e^{- i k t} J_{\l + \frac{1}{2} } (| k| r ) .
\end{equation}
From the definition of the canonical conjugate momentum in (\ref{ccm}) , the corresponding 
Fourier mode is
\begin{equation} \label{amsol2}
\pi_{( \l, m ) }^{\mm \, *} =  - i k a_{\mm (\l, m) } (k) r^{-\frac{1}{2}}  e^{- i k t} J_{\l + \frac{1}{2} } (|k| r ) .
\end{equation}
Using the solutions in (\ref{amsol1}) and (\ref{amsol2})  we can write the mode expansion as 
 \begin{eqnarray}\label{modexp2}
 A_{\mm ( \l. m) } ( r, t ) &=& \frac{1}{\sqrt{2} }  
 \int_0^\infty  dk \left( a_{\mm (\l, m) } (k) e^{ - i k t} +  (-1)^m
 a^\dagger_{\mm (\l, -m )}(k) e^{i k t} 
  \right) r^{-\frac{1}{2}}  J_{\l + \frac{1}{2} } ( k r) , \\ \nonumber
 \pi^{\mm\, *}_{( \l, m ) } ( r, t) &=&   \frac{ 1 }{\sqrt{2} } 
 \int_0^\infty  k  dk \left( - i a_{\mm (\l, m) } (k)  
 ( k r) e^{ - i k t} + i (-1)^m a^\dagger_{\mm (\l, -m )}(k) e^{i k t} 
   \right)  r^{-\frac{1}{2}} 
  J_{\l + \frac{1}{2} } (kr) .
 \end{eqnarray}
 We also have 
 \begin{eqnarray}
 \pi^{\mm}_{( \l, m ) } ( r, t) &=&   \frac{ 1 }{\sqrt{2} } 
 \int_0^\infty  k  dk \left(  i a_{\mm (\l, m) }^\dagger  (k)  
  e^{  i k t} -i (-1)^m a_{\mm (\l, -m )}(k) e^{-i k t} 
   \right)  r^{-\frac{1}{2}} 
  J_{\l + \frac{1}{2} } (kr). \nonumber \\
 \end{eqnarray}
 We can promote the fields $A_\mm, \pi^\mm$ to operators by promoting 
 $a_{\mm (\l, m ) }( k) ,  a^\dagger_{\mm (\l, m ) }( k)$ to operators. 
 Then the  canonical commutation relations  
 \begin{equation}
 [A_{\mm ( \l. m) } ( r, t ) , \pi^\mm_{( \l', m') } ( r', t)  = i \delta ( r- r') \delta_{\l, \l'} \delta_{m, m'} .
 \end{equation}
 imply that the commutation relations 
 \begin{equation} \label{crean2}
 [a_{\mm (\l, m )} (k) , a^{\dagger}_{\mm (\l', m')} (k') ] = \delta( k-k') \delta_{\l, \l'} \delta_{m, m'}.
 \end{equation}

\subsubsection*{Electric correlator on the sphere and edge entanglement}

We  proceed to evaluate the two-point function of the 
radial component of the electric field. 
Examining the components of the field strengths in (\ref{fieldstr}) and using the definition of the canonical 
momenta (\ref{ccm}), 
we see that the electric field $F_{\hat t \hat r}$ is related 
to the momentum $\pi^r$
\begin{eqnarray}
F_{\hat t \hat r} &=& \sum_{\ell,m}(\dot{A}_{r; \ell m}-\partial_r A_{0; \ell m})Y_{\ell m} (\theta, \phi) , \\ \nonumber
&=& \sum_{\l, m } \frac{\pi^{r\, *} _{(\l, m ) } (r, t) }{r^2 } Y_{\ell m } (\theta, \phi) .
\end{eqnarray}
Using the mode expansion in (\ref{modexp1}), 
we can proceed to evaluate the two-point function of the electric field
\begin{eqnarray}
&& \langle 0| F_{\hat t \hat r} (t,  r,  \theta, \phi)   F_{\hat t' \hat r'} ( t, r',  \theta', \phi ') |0\rangle  = 
\\ \nonumber
&& \frac{1}{ 2 ( rr')^{\frac{3}{2} } }  \sum_{\l, \l', m, m'}  \l ( \l +1) \l' ( \l' +1)  \int_0^\infty dk dk' 
\left[ 
J_{\l + \frac{1}{2} } ( k r)  J_{\l' + \frac{1}{2} } ( k' r')    \right. \\ \nonumber
& & \qquad\qquad\qquad\qquad \left. 
 \times  (-1)^{m'} 
 \langle 0| a_{ r (\l, m ) }  (k) a^\dagger_{ r ( \l', m' ) }(k') |0\rangle Y_{\l, m } ( \theta, \phi) Y_{\l', - m'} (\theta', \phi')
 \right].
\end{eqnarray}
Using the commutation relations in (\ref{crean1}), we obtain
\begin{eqnarray}
 && \langle 0| F_{\hat t \hat r} (t,  r,  \theta, \phi)   F_{\hat t' \hat r'} ( t, r',  \theta', \phi ') |0\rangle  = \\ \nonumber
 && \qquad\qquad
 \frac{1}{ 2 ( rr')^{\frac{3}{2} } }\sum_{\l, m } \l ( \l +1) \int_0^\infty  dk  J_{\l + \frac{1}{2} } ( k r)  J_{\l' + \frac{1}{2} } ( k' r')  
 Y_{\l, m } ( \theta, \phi) Y_{\l,  m}^* (\theta', \phi'). \nonumber 
\end{eqnarray}
We perform the integral using the identity in  \cite{gradshteyn2007}, see   page 696 equation 3 of {\bf 6.612}
\begin{align}\label{besselidentity}
   \int_0^\infty dk  J_{\ell+\frac{1}{2}}(kr)J_{\ell+\frac{1}{2}}(kr')=\frac{1}{\pi\sqrt{rr'}}Q_{\ell}(\frac{r^2+r'^2}{2rr'}).
\end{align}
Where $Q_{\nu}(z)$ is the Legendre function of second kind which can be written in terms of 
the hypergeometric function \cite{gradshteyn2007}, see   page 1024 equation 2 of {\bf 8.820}
\begin{align}
    Q_{\nu}(z)=\frac{\Gamma \left(\frac{1}{2}\right) \Gamma (\nu +1) z^{-\nu -1} \, _2F_1\left(\frac{\nu +2}{2},\frac{\nu +1}{2};\frac{1}{2} (2 \nu +3);\frac{1}{z^2}\right)}{2^{\nu +1} \Gamma \left(\nu +\frac{3}{2}\right)}.
\end{align}
Therefore the corrrelator is given by 
\begin{equation}\label{eleccor}
 \langle 0| F_{\hat t \hat r} (t,  r,  \theta, \phi)   F_{\hat t' \hat r'} ( t, r',  \theta', \phi ') |0\rangle  
 =\sum_{\ell,m} \frac{\l(\l+1) }{2\pi  (rr')^2} Q_{\ell}  (\frac{r^2+r'^2}{2rr'})
 Y_{\l, m } (\theta, \phi) Y_{\l, m }^*(\theta', \phi') .
\end{equation}
In equation (A.12) of \cite{Soni:2016ogt},  
this correlator was evaluated in the Coloumb gauge.
The authors evaluated the electric field correlator 
in Cartesian coordinates, transformed to polar coordinates and then decomposed the 
answer in spherical harmonics. 
The answer for each harmonic was not obtained in closed form unlike the result in 
(\ref{eleccor}). In the appendix \ref{appen}, we compare our result with their result and show that our closed 
form result for each $\l$ agrees with the expansion found in \cite{Soni:2016ogt}.

From the discussion in section (\ref{sec2.1}), we see that we need the 2-point functions on the sphere,
in the limit that the two-points have the same radial position. 
Therefore let us take
\begin{equation}
r' = r + \delta,  \qquad \delta \rightarrow 0.
\end{equation}
In this limit, the Legendre function of the second kind can be expanded as 
\begin{equation}\label{Qexpand}
\lim_{\delta\rightarrow 0} Q_\l \left( \frac{ r^2 + ( r+\delta)^2}{ 2 r ( r+\delta)} \right) 
= \frac{1}{2 } \log( \frac{ r^2}{\delta^2} )  + \log 2 - H_\l + O(\delta) .
\end{equation}
where $H_\l$ refers to the Harmonic number. 
Note that the leading divergence is independent of $\l$. 
Substituting this limit in the correlator (\ref{eleccor}), we obtain 
\begin{equation}
\lim_{\delta\rightarrow 0} G_{rr} ( r, r+\delta; x, y) =  \frac{1}{4\pi r^4}  \log ( \frac{r^2}{\delta^2} )  \sum_{\l \geq 1, m }
\l(\l +1) Y_{\l, m } ( \theta, \phi) Y_{\l, m }^*( \theta'\, \phi'),
\end{equation}
where we retain only the leading term in the $\delta\rightarrow 0$ limit. 
Thus the correlator is diagonal in the angular momentum basis and the diagonal elements are 
given by 
\begin{equation}
\lim_{\delta\rightarrow 0} \langle \l, m |G_{rr} ( r, r+\delta) |l' m' \rangle  = \frac{\l(\l +1)}{4\pi r^4} \log ( \frac{r^2}{\delta^2} )  \delta_{\l, \l'} \delta_{m, m'} .
\end{equation}
As mentioned earlier, we see that the correlator is proportional to the Laplacian of 
the massless scalar  on $S^2$. 
From (\ref{unived}), we see  that 
the edge mode contribution to the entanglement  is obtained by evaluating the  log-determinant of this
operator.  The coefficient  which is proportional to the $\log( R/\epsilon)$ where $R$ is the radius of the 
entangling sphere  and $\epsilon$ is  a cutoff on the sphere, can be obtained from 
\begin{equation}
S_{\rm edge} ( \rho_A ) =  \frac{1}{2} \sum_{\l=1}^\infty ( 2\l +1) \log\big[\l ( \l +1) \big].
\end{equation}
Here we are ignoring all terms which are proportional which grow as the area $R^2/\epsilon^2$ and 
retained the term which contains the  logarithmic coefficient to the entanglement entropy. 

We can use the standard methods to evaluate the determinant of the scalar Laplacian to obtain the 
 contribution to the entanglement entropy from the edge modes. 
The free energy of the of the massless scalar or the 0-form is given by
\begin{equation}
-\frac{1}{2} \log ({\rm det} \Delta_0^{ S^2})  = -\frac{1}{2} \sum_{\l =1}^\infty ( 2\l +1) \log ( \l ( \l +1) ).
\end{equation}
Then using the methods of \cite{Anninos:2020hfj,David:2021wrw} we  can write the Harish-Chandra character integral representation
of the free energy. This is given by 
\begin{equation}
-\frac{1}{2} \log {\rm det} ( \Delta_0^{S^2} )  = 
\int_0^\infty \frac{dt}{2t} \frac{ 1 + e^{-t}}{ 1- e^{-t} } \chi^{dS}_{(1, 0)} (t).
\end{equation}
where $\chi^{dS}_{(2, 0)} (t)$ is the $SO(1, 2) $ Harish-Chandra character of the $0$-form in the 
$\Delta = \frac{1}{2} + i \nu$ representation with $i\nu = \frac{1}{2}$. 
This character is given by
\begin{equation}
\chi^{dS}_{(1, 0)} (t) = \frac{ 1 + e^{-t}}{  1- e^{-t} } .
\end{equation}
Therefore, the result for the free energy is 
\begin{equation}\label{charrep}
-\frac{1}{2} \log {\rm det} ( \Delta_0^{S^2} ) =
 \int_0^\infty \frac{dt}{2t} \frac{ 1 + e^{-t}}{ 1- e^{-t} }   \frac{ 1 + e^{-t}}{  1- e^{-t} } .
 \end{equation}
 In the character integral representation the coefficient of the $\log ( \frac{R}{\epsilon}) $ 
 term to the free energy is easy to read out. 
 It is given by the coefficient of the $t^{-1}$ term of the integrand in (\ref{charrep}) which is $\frac{1}{3}$. 
 Therefore, the  logarithmic contribution of the edge modes to the entanglement entropy is given
 by 
 \begin{eqnarray}
S_{\rm edge} ( \rho_A ) &=&   \frac{1}{2} \log {\rm det} ( \Delta_0^{S^2} ) , \\ \nonumber
&=& -\frac{1}{3} \log \frac{R}{\epsilon} .
\end{eqnarray}
We see that  this result agrees with that evaluated in \cite{Soni:2016ogt}.  Here we emphasize that this entanglement 
is the contribution of the electric centre of the local alegbra as discussed in \cite{Casini:2013rba}. 
This resulted from our choice of labelling the superselection sectors using the electric Gauss law.

\subsubsection*{Magnetic correlator on the sphere and edge entanglement}

As discussed in \cite{Casini:2013rba}, the entanglement entropy of the centre of the algebra depends on the 
choice centre. We can also consider the 
 the case where the superselection sectors 
are labelled by the magnetic field, the magnetic centre.
The magnetic field satisfies the condition
\begin{equation}
\nabla^i B_i = 0.
\end{equation}
Following the same arguments as that for the electric field, this implies that 
the radial component of the $B_{\hat r}$ must agree across the entangling surface. 
Therefore the field strength $F_{\hat e \hat m} = B_{\hat r} $ labels superselection sectors. 
The arguments in section (\ref{sec2.1}) , then imply that 
we would need  the
two-point function of the magnetic field on the sphere to evaluate the entanglement
entropy of the edge modes. 

From (\ref{fieldstr}), we see, that the magnetic flux is given by 
\begin{equation}
F_{\hat e \hat m} = \sum_{\l, m } \frac{\sqrt{ \l ( \l + 1) }}{r} A_{\mm; \l m } Y_{\l m } .
\end{equation}
Using the mode expansion in (\ref{modexp2}), the two-point function of the magnetic field is 
given by 
\begin{eqnarray}
&& \langle 0| F_{\hat e \hat m} (t,  r,  \theta, \phi)   F_{\hat e' \hat m'} ( t, r',  \theta', \phi ') |0\rangle  = 
\\ \nonumber
&&  \qquad\qquad
\frac{1}{ 2 ( rr')^{\frac{3}{2} } }  \sum_{\l, \l', m, m'}  \sqrt{ \l ( \l+1) \l'( \l' +1) } 
 \int_0^\infty dk dk' 
\left[ 
J_{\l + \frac{1}{2} } ( k r)  J_{\l' + \frac{1}{2} } ( k' r')    \right.  \nonumber \\ \nonumber
 & &\qquad\qquad\qquad\qquad \left. \times  (-1)^{m'}
 \big\langle 0| a_{ \mm (\l, m ) }  (k) a^\dagger_{ \mm ( \l', m' ) }(k') |0\big\rangle \; Y_{\l, m } ( \theta, \phi) Y_{\l', - m'} (\theta', \phi')
 \right].
\end{eqnarray}
Using the commutation relations in (\ref{crean2})  , we get
\begin{eqnarray}
&& \langle 0| F_{\hat e \hat m} (t,  r,  \theta, \phi)   F_{\hat e' \hat m'} ( t, r',  \theta', \phi ') |0\rangle  = \\ \nonumber
&& \qquad\qquad \frac{1}{ 2 ( rr')^{\frac{3}{2} } }\sum_{\l, m } \l ( \l +1) \int_0^\infty  dk  J_{\l + \frac{1}{2} } ( k r)  J_{\l' + \frac{1}{2} } ( k' r')  
 Y_{\l, m } ( \theta, \phi) Y_{\l,  m}^* (\theta', \phi') .
\end{eqnarray}
Then using the identity (\ref{besselidentity}), we obtain 
\begin{equation}\label{magccor}
 \langle 0| F_{\hat e \hat m} (t,  r,  \theta, \phi)   F_{\hat e' \hat m'} ( t, r',  \theta', \phi ') |0\rangle 
 =\sum_{\ell,m} \frac{\l(\l+1) }{2\pi  (rr')^2} Q_{\ell}  (\frac{r^2+r'^2}{2rr'})
 Y_{\l, m } (\theta, \phi) Y_{\l, m }^*(\theta', \phi') .
\end{equation}
Comparing (\ref{eleccor}) and ( \ref{magccor}) we see that they are identical. 
Following the same steps as in the case of the electric superselection sectors,  we see that the 
logarithmic contribution to 
entanglement entropy when  the magnetic field labels superselection sectors is 
given by 
 \begin{eqnarray}
S_{\rm edge} ( \rho_A )|_{\rm magnetic} &=&   \frac{1}{2} \log {\rm det} ( \Delta_0^{S^2} ) , \\ \nonumber
&=& -\frac{1}{3} \log \frac{R}{\epsilon} .
\end{eqnarray}
Thus the entanglement entropy of the magnetic centre coincides with that of the electric centre. 
This is due to the electric magnetic duality of the $U(1)$ theory. 
In general choice of different centres can result in different entanglement associated with the centre. 
In \cite{Casini:2013rba} it was shown in lattice gauge theory  one could choose a trivial centre resulting no 
entanglement entropy of the centre.

\subsection{$U(1)$ theory in arbitrary even $d$ } \label{u1arbd}

In this section we generalise the discussion to arbitrary even $d$.
Let us expand the gauge potential as
\begin{eqnarray}
A_\mu &=& \sum_{\l, \lambda, m} \left( 
A_{0 ( \l , \lambda, m )} ( r, t) Y_{\l \lambda m, \mu  }^{{\bf 0}} 
+  A_{r ( \l , \lambda, m )} ( r, t) Y_{\l \lambda m, \mu  }^{{\bf r}}  \right.  \\ \nonumber
&& \qquad \left. +  A_{e ( \l , \lambda, m )} ( r, t) \hat Y_{\l \lambda m, \mu  }^{{\bf e}} 
+
 A_{ \vec\mm ( \l , \lambda, m )} ( r, t) Y_{\l \lambda m, \mu  }^{{\vec \mm}}  \right) .
\end{eqnarray}
where the  $d$ dimensional covariant vectors are defined as
\begin{eqnarray}
Y_{\l \lambda m   }^{{\bf 0}}  = \{ Y_{\l \lambda m }(\Omega ) , 0 , \cdots 0 \}, \\ \nonumber
Y^{\bf r}_{\l \lambda m} = \{ 0, Y_{\l \lambda m}( \Omega ), 0, \cdots 0 \} , \\ \nonumber
 Y_{\l \lambda m }^{{\bf e}}  = \frac{r}{\sqrt{\l ( \l  + \hat d -1) } }
 \{ 0, 0,  \partial_i  Y_{\l \lambda m}( \Omega )\}, \\ \nonumber
 \hat Y_{\l \lambda m }^{{\vec \mm}}  = \{ 0, 0, Y_{\l \lambda m }^ {\vec \mm} ( \Omega)   \}.
\end{eqnarray}
$Y_{\l  \lambda m }$ are scalar spherical harmonics on $S^{\hat d}$, $l$ is the principal quantum number, 
$m$ the azimuthal quantum number and $\lambda$ refers to the rest of the $\hat d -2$ quantum numbers. 
It is important to note that the azimuthal quantum number can take values both in the positive and 
negative values in the set of integers. 
These harmonics can be found in \cite{Higuchi:1986wu}. The principal quantum number takes values $\l =0, 1, \cdots$. 
The variable $\Omega $ refers to all the angles on the sphere $S^{\hat d}$. 
We work with the metric 
\begin{equation}
ds^2 = - dt^2 + dr^2 + r^2 d\Omega^2.
\end{equation}
where the metric on sphere is the conventional round metric. 
The partial derivative in the definition of $ Y_{\l \lambda m }^{{\bf e}} $ refers to derivative in the 
angular directions of the sphere. 
Finally $Y_{\l \lambda m }^ {\vec \mm} ( \Omega) $ refers to the other $\hat d-1$ vector harmonics on 
$S^{\hat d}$.  The properties of both the scalar and vector harmonics can be found in \cite{Higuchi:1986wu}. 
From the construction\footnote{The scalar harmonics are products of associated Legendre functions 
of the first kind and a phase. The phase and one of the associated Legendre functions can be grouped 
to form the spherical harmonic on $S^2$. The property of these functions under conjugations can be 
obtained from this observation.}
of these harmonics we see that they obey the property 
\begin{equation}
Y_{\l \lambda m }( \Omega)^* = (-1)^m  Y_{\l \lambda - m }( \Omega).
\end{equation}
Then the reality of the vector potential implies that we have the relations 
\begin{eqnarray} \label{realityd}
A_{0 ( \l , \lambda, m )}^* ( r, t)  = (-1)^m A_{0 ( \l , \lambda, m )} ( r, t) ,  \\ \nonumber
A_{r ( \l , \lambda, m )}^* ( r, t) = (-1)^m A_{r ( \l , \lambda, m )} ( r, t).
\end{eqnarray}
Following the same discussion as in the case of 4d, we can fix gauge so that $A_{e} =0$. 
It is sufficient to focus on the  components $A_0$ and $A_r$ to obtain the 
entanglement entropy of the superselection sectors corresponding to the electric field. 
The action for these components derived from the Maxwell action  is given by 
\begin{eqnarray}
S &=& \sum_{l \lambda m =-\l}^\l \int dr L_{\l \lambda m } ,\\ \nonumber
L_{\l \lambda m } &=& \frac{1}{2}\left[  r^{\hat d} \dot A_{r (\l , \lambda, m } \dot  A^*_{r ( \l, \lambda, m ) }
+ r^{\hat d} \partial_r A_{0 ( \l, \lambda, m ) } \partial_r A_{0 ( \l, \lambda m ) }^* \right. \\ \nonumber
&& \left. - r^{\hat d} \dot A_{r ( \l, \lambda, m )} \partial_r A^*_{0(\l ,\lambda, m) } 
- r^{\hat d} \dot A^*_{r ( \l, \lambda m ) } \partial_r A_{0 ( \l, \lambda, m ) }
+ \l ( \l + \hat d - 1)  r^{\hat d-2} A_{0 ( \l \lambda m ) }  A_{0 ( \l ,\lambda ,m ) } ^* \right].
\end{eqnarray}
The canonical conjugate momentum to $A_r$ is given by 
\begin{equation} \label{pirdef}
\pi^{r\, *} _{( \l,  \lambda , m ) } = r^{\hat d} ( \dot A_{r ( \l, \lambda, m)} - \partial_r A_{0 ( \l, \lambda, m ) } ) . 
\end{equation}
Using the constraint for $A_0$, we can write
\begin{equation}
\partial_r \pi^{r\, *}_{( \l, \lambda, m )} + \l ( \l  + \hat d - 1) A_{0 ( \l, \lambda, m ) }=0. 
\end{equation}
Eliminating $A_0$, we obtain the equation of motion for  the $k$-th Fourier mode $A_r$, which is given by 
\begin{eqnarray}
\partial_r^2  A_{r ( \l, \lambda, m ) } + \frac{ \hat d -2}{r} \partial_r  A_{r ( \l, \lambda, m ) }
+ \Big(  k^2 - \frac{ \l ( \l + \hat d -1) + \hat d-2}{r^2} \Big)  A_{r ( \l, \lambda, m ) } =0.
\end{eqnarray}
The solution which is regular at the origin is given by 
\begin{equation}\label{solnard}
A_{r ( \l, \lambda, m ) }  =  e^{-i k t} a_{r ( \l, \lambda, m ) } 
  r^{ - \frac{\hat d - 3}{2}} J_{\l + \frac{ \hat d -1}{2} } ( |k| r) ,
\end{equation}
where $a_{r ( \l, \lambda, m ) } (k)  $ is the integration constant. 
The equation for the momenta can be obtained by using the definition in (\ref{pirdef}), eliminating 
$A_0$ and substituting for $A_r$ from (\ref{solnard}). This results in 
\begin{eqnarray}
\pi^{r\, *}_{ (\l, \lambda, m ) } = r^{\hat d} 
\left( - i k  e^{-i k t} a_{r ( \l, \lambda, m ) } (k)  r^{ - \frac{\hat d - 3}{2}} J_{\l + \frac{ \hat d -1}{2} } ( |k| r) 
+
\frac{1}{ \l ( \l + \hat d -1) }   \partial_r
\Big(  r^{\hat d -2} \partial_r \pi^{r\, *}_{ (\l, \lambda, m ) }\Big) \right) .  \nonumber \\
\end{eqnarray}
The solution to this differential equation is given by 
\begin{eqnarray}
&&\pi^{r\, *}_{ (\l, \lambda, m ) }  = c_1r^{\hat d - 1+\l } + c_2 r^{-\l}   \\ \nonumber
&& +a_{r ( \l, \lambda, m ) } (k) e^{-i k t}  \left(  \frac{i 2^{- ( \l  + \frac{\hat d - 1}{2} )} 
\l ( \l + \hat d -1)  k^{\l + \frac{\hat d -3}{2}} }{ \Gamma( \l + \frac{ \hat d +1}{2} ) } r^{ \l + \hat d +1} 
- i \frac{\l ( \l + \hat d - 1)  r^{\frac{ \hat d -1}{2}} }{k }  J_{\l + \frac{\hat d -1}{2}} ( |k| r) 
\right).
\end{eqnarray}
 $\l \geq 1$ for these modes, regularity at the origin results in $c_2 =0$ and demanding that the solution is 
 well behaved  at infinity determines $c_1$ giving
 \begin{eqnarray}\label{solndpi}
 \pi^{r\, *}_{ (\l, \lambda, m ) }  =  - i \frac{\l ( \l + \hat d - 1)  }{k } 
 a_{r ( \l, \lambda, m ) } (k) e^{-i k t}     r^{\frac{ \hat d -1}{2}}
 J_{\l + \frac{\hat d -1}{2}} ( |k| r) .
 \end{eqnarray}
 
 Using the classical solutions (\ref{solnard}) and (\ref{solndpi}) for each Fourier mode, we can 
 write down the mode expansions
\begin{eqnarray}\label{modexpd1}
 A_{r ( \l, \lambda, m) } ( r, t ) &=& \frac{1}{\sqrt{2} } 
 \int_0^\infty 
  k dk \left( a_{r (\l,  \lambda, m) } (k) e^{ - i k t}  + (-1)^m  a_{r( \l, \lambda, -m )}^\dagger (k)  e^{i k t}  \right) 
  r^{- \frac{\hat d -3}{2} }  J_{\l + \frac{\hat d -1 }{2} } ( k r) ,  \nonumber \\ \nonumber
 \pi^{r\, *}_{( \l,\lambda,  m ) } ( r, t) &=&   \frac{ \l ( \l +\hat d -1 ) }{\sqrt{2} } 
 \int_0^\infty  dk \left( - i a_{r (\l, \lambda, m) } (k) e^{ - i k t} +  i (-1)^m a_{r(\l, \lambda,  -m)}^\dagger(k)  e^{i k t}    \right)
 r^{\frac{\hat d-1}{2}}
 J_{\l + \frac{\hat{d}-1}{2} } ( k r) . \\
 \end{eqnarray}
 This  form for the mode expansion respects the reality condition (\ref{realityd}).
 Let us also write the mode expansion 
 \begin{eqnarray}\label{modexpd2}
 \pi^{r}_{( \l, m ) } ( r, t) &=&   \frac{ \l ( \l +\hat d -1 ) }{\sqrt{2} } 
  \int_0^\infty  dk \left(  i a_{r (\l, \lambda, m) }^\dagger (k) e^{  i k t} -  i (-1)^m a_{r(\l,  \lambda, -m)}(k) e^{-i k t} 
    \right) r^{\frac{\hat d -1}{2}} J_{\l + \frac{1}{2} } ( k r) . \nonumber \\
 \end{eqnarray}
The equal time commutation relation of the fields $A_r, \pi^r$ is given by 
\begin{equation}
[A_{r (\l, \lambda, m ) } (r, t) , \pi^r_{r ( \l', \lambda', m' ) } ( r, t) ] = i \delta_{\l, \l'} \delta_{\lambda, \lambda'}\delta_{m, m'} \delta( r- r') .
\end{equation}
These relations together with the mode expansions (\ref{modexpd1}), ( \ref{modexpd2})  imply the 
following commutation relations between the creation and annihilation operators
\begin{equation}
[a_{r(\l, \lambda, m ) } (k) , a^\dagger_{r(\l', \lambda', m' ) (k')  }] = \frac{1}{\l ( \l + \hat d - 1) } \delta_{\l \l'} \delta_{\lambda, \lambda'}\delta_{m, m'}.
\end{equation}

\subsubsection*{Electric field correlators on $S^{\hat d}$ }

The electric field is related to  the canonical momentum  by 
\begin{eqnarray}
F_{\hat t \hat r } &=& \sum_{\l, \lambda, m } \Big ( \dot A_{r ( \l, \lambda, m )} - \partial_r A_{0 ( \l, \lambda, m ) } 
\Big) Y_{\l\lambda m } ( \Omega)  , \\ \nonumber
&=& \sum_{\l, \lambda, m }  \frac{ \pi^{r\, *}_{ ( \l, \lambda, m ) } }{ r^{\hat d } }  Y_{\l\lambda m } ( \Omega) .
\end{eqnarray}
where we have used (\ref{pirdef}). 
Proceeding as in the case of $d=4$, we evaluate the two-point function of the radial electric field 
\begin{eqnarray}\label{finalu1cor}
 \langle 0| F_{\hat t \hat r} (t,  r,  \Omega)    F_{\hat t' \hat r'} ( t, r', \Omega' ) |0\rangle  
 =\sum_{\ell, \lambda, m} \frac{\l(\l+\hat d - 1 ) }{2\pi  (rr')^{\frac{d}{2} }} Q_{\ell}  (\frac{r^2+r'^2}{2rr'})
 Y_{\l, \lambda, m } (\Omega ) Y_{\l, \lambda, m }^*(\Omega' ) . \nonumber \\
\end{eqnarray}
The result is a generalisation of that seen for the case of $d=4$ in (\ref{eleccor}), we see that the correlator is proportional 
to the Laplacian of the scalar or the 0-form on the sphere $S^{\hat d }$. 
We can take the 
coincident limit in the radial direction. From the leading divergence 
we evaluate the contribution of the edge modes to the entanglement entropy. 
The logarithmic coefficient is given by 
\begin{eqnarray}
S_{\rm edge} ( \rho_A) = \frac{1}{2} \sum_{l =1}^\infty g_{\l, d} \log ( \l ( \l + \hat d -1)) ,  \qquad
g_{\l, d} = \frac{  ( 2\l + d - 3) \Gamma( \l + d - 3) }{ \l! \Gamma( d-2) } ,
\end{eqnarray}
where $g_{\l, d}$ are the degeneracies of the Laplacian  of the 0-form on $S^{\hat d}$. 
Note that the $\l=0$ mode is not counted, and therefore  the logarithmic coefficient 
entanglement entropy of the 
edge modes coincides precisely with that of the 
the free energy of the $0$-form. 
Using the methods of \cite{Anninos:2020hfj,David:2021wrw}, 
the partition function of the $0$-form can be written in terms of its 
Harish-Chandra characters
\begin{equation}
-\frac{1}{2} \log \rm{det} (\Delta^{S^{\hat d}}_0 ) = \int_0^\infty\frac{dt}{2t} \frac{1 + e^{-t}}{ 1- e^{-t}} 
\chi_{\hat d, 0}.
\end{equation}
Here $\chi_{\hat d, 0}$ is the  $SO(\hat d, 1 )$ Harish-Chandra character of the 0-form in the 
$\Delta = \frac{\hat d -1}{2} + i \nu $ representation with $i\nu = \frac{\hat d -1}{2} $ which is given by 
\begin{equation}
\chi_{\hat d, 0}^{\rm dS}(t) = \frac{1 + e^{- ( \hat d -1 ) t} }{ 1- e^{-( \hat d - 1) t}} .
\end{equation}
Therefore the logarithmic coefficient of the entanglement entropy is obtained from the 
$1/t$ coefficient of the integrand in 
\begin{eqnarray} \label{sedged}
S_{\rm edge} ( \rho_A) &=& \frac{1}{2} \log \rm{det} (\Delta^{S^{\hat d}}_0 ) , \\ \nonumber
&=& - \int_0^\infty\frac{dt}{2t} \frac{1 + e^{-t}}{ 1- e^{-t}}  
\frac{1 + e^{- ( \hat d -1 ) t} }{ 1- e^{-( \hat d - 1) t}}. 
\end{eqnarray}

\subsection{The $U(1)$ theory on spheres and entanglement of edge modes}

In this section we consider the representation of the partition function 
the $U(1)$ theory on the  even $d$ dimensional sphere in terms of the Harish-Chandra 
character as constructed in \cite{Anninos:2020hfj,David:2021wrw}. 
We show that the contribution to the partition function from the 
edge character coincides with the contribution of the superselection sectors to the entanglement 
entropy. 
After gauge fixing, the partition function of the $U(1)$ theory on $S^d$ can be written as 
\begin{eqnarray}
\log {\cal Z}_1 [S^{d}] = - \frac{1}{2} \log( \rm{det}_T\Delta_1^{S^{d}} )
+ \frac{1}{2} \log( \rm{det}' \Delta_0^{S^{d}} ).
\end{eqnarray}
Here $\Delta_1$ is the Laplacian on $1$-forms and the subscript $T$ refers to the fact that the 
determinant is evaluated  over transverse $1$-forms.  The prime in the second term is to denote
that the determinant ignores the zero mode of the scalar. 
In \cite{David:2021wrw} it was shown that the partition function of $p$-forms can be written in 
terms of Harish-Chandra characters. Using these results we get
\begin{eqnarray}
- \frac{1}{2} \log({ \rm{det}}_T\Delta_1^{S^{ d}} ) &=& 
\int_0^\infty \frac{dt}{2t} \frac{ 1+ e^{-t}}{ 1-e^{-t}} \left( \chi^{dS}_{( d-1, 1)} (t)  - \chi^{dS}_{(d-3, 0) }(t) \right) , 
\\ \nonumber
 \chi^{dS}_{( d-1, 1)} (t) & =& (d-1) \frac{ e^{-(d-2) t} + e^{-t} }{ ( 1- e^{-t} )^{d-1} } ,  \qquad\qquad
 \chi^{dS}_{( d-3, 0)} (t)  =   \frac{ e^{-(d-3) t} + 1 }{ ( 1- e^{-t} )^{d-3} }, \\  \nonumber
 -\frac{1}{2}  \log({ \rm{det}'} \Delta_0^{S^{d}} ) &=&  \int_0^\infty \frac{dt}{2t} \frac{ 1+ e^{-t}}{ 1-e^{-t}} \chi^{dS}_{(d-1, 0)}(t) , \qquad\quad
  \chi^{dS}_{( d-1, 0)} (t)  = \frac{ e^{-( d-1) t} +1}{ ( 1- e^{-t} )^{d-1} }.
\end{eqnarray}
From this representation of  the one loop partition function of the $1$-form, we see that 
it can be written as a contribution from  bulk and edge characters as 
\begin{eqnarray}
- \frac{1}{2} \log( \rm{det}_T\Delta_1^{S^{ d}} ) &=&  \int_0^\infty \frac{dt}{2t} \frac{ 1+ e^{-t}}{ 1-e^{-t}} 
( \chi_{\rm bulk} (t)  - \chi_{\rm edge} (t) ) , \\ \nonumber
\chi_{\rm bulk} (t)  &=&   \chi^{dS}_{( d-1, 1)} (t)  -    \chi^{dS}_{( d-1, 0)} (t), \\ \nonumber
\chi_{\rm edge} (t) &=&  \chi^{dS}_{( d-3, 0)} (t) .
\end{eqnarray}
Note that the edge character is the Harish-Chandra character for the $0$-form on a sphere of 
 $d-2 =\hat d$ dimensions. 
Comparison with (\ref{sedged}), we see that its contribution to the logarithmic coefficient of the 
free energy of the $U(1)$ theory on $S^d$ 
including the sign agrees precisely 
with that the of the entanglement entanglement entropy of the edge modes.

\section{Graviton in $d=4$ dimension} \label{gravsection}

In this section we evaluate the contribution of the edge modes to the 
logarithmic coefficient of the 
entanglement entropy of gravitons across a spherical surface in $d=4$. 
Our discussion will closely follow that of the $U(1)$ theory. 
The  Lagrangian for the theory of linearized gravitons is given by 
\begin{eqnarray} \label{laggrav}
{\cal L } = - \partial_\mu h^{\mu\nu} \partial_\alpha h^\alpha_{\, \nu} 
+ \frac{1}{2} \partial^\alpha h_{\mu\nu} \partial_\alpha h^{\mu\nu} 
+ \partial_\mu h^{\mu \nu} \partial_\nu h^\alpha_{\, \alpha}
- \frac{1}{2} \partial_\alpha h^{\mu}_{\, \mu} \partial^\alpha h^\nu_{\, \nu}.
\end{eqnarray}
This Lagrangian admits the gauge symmetry
\begin{equation}\label{gsym}
\delta h_{\mu\nu}   = \partial_\mu \xi_\nu + \partial_\nu \xi_\mu.
\end{equation}
In the linearized theory, the curvature is gauge-invariant. It is given by 
\begin{equation} \label{reim}
R_{\mu\nu\rho\sigma} = \frac{1}{2} ( \partial_\nu \partial_\rho h_{\mu \sigma}
-\partial_\mu \partial_\rho h_{\nu \sigma}
+\partial_\mu \partial_\sigma h_{\nu \rho} - \partial_\nu \partial_\sigma h_{\mu \rho} ). 
\end{equation}
The equations of motion from the action (\ref{laggrav}) are the Einstein equations
\begin{equation}
R_{\mu\nu} - \frac{1}{2} \eta_{\mu\nu} R=0.
\end{equation}

In the case of the Maxwell field, the Gauss law resulted
in superselection sectors.  To obtain a gauge-invariant characterization of the superselection 
sectors in gravity it is convenient to think of 
the Riemann curvature (\ref{reim}) as a field strength of a $U(1)$ gauge 
potential. 
Consider the  gauge potential constructed out of the graviton
\begin{eqnarray}
A_{\mu\alpha\beta} = \frac{1}{2} (  \partial_\beta h_{\mu \alpha} - \partial_\alpha h_{\mu\beta} ) .
\end{eqnarray}
Under the gauge symmetry (\ref{gsym}) the potential transforms as an $U(1)$ gauge field
\begin{eqnarray}
\delta A_{\mu\alpha\beta}  = \partial_\mu \lambda_{\alpha\beta}, \qquad
\lambda_{\alpha\beta} = \frac{1}{2} ( \partial_{\beta} \xi_\alpha - \partial_\alpha\xi_\beta) .
\end{eqnarray}
The Riemann curvature can then be written as as a field strength of this $U(1)$ gauge field. 
\begin{eqnarray}
R_{\mu\nu\alpha\beta} =  \partial_{\mu } A_{\nu \alpha \beta} - \partial_{\nu} A_{\mu \alpha\beta} .
\end{eqnarray}
Indeed in \cite{Casini:2021vmi},   the components $R_{0i 0j}$  where $i, j$ are spatial indices were 
identified as the electric fields while  $B_{ij} = \frac{1}{2} \epsilon_{ilm} R^{lm}_{\;\;0j}$ were 
identified as magnetic fields.  

We show that the Riemann curvature  $R_{0\mu \alpha\beta}$ obeys a similar Gauss law 
as that in electrodynamics once the Hamiltonian constraint together with the on shell conditions are satisfied. 
The Gauss constraint or the Hamiltonian constraint  in gravity  are the following
\begin{eqnarray}
R_{0i} &=&0 \label{g1}, \\ 
R_{00} + \frac{1}{2} R &=&0 \label{g2}.
\end{eqnarray}
Consider the Bianchi identity in linearised gravity,
\begin{equation}
\partial_\lambda R_{\mu \nu \alpha\beta} + \partial_\mu R_{\nu \lambda\alpha\beta} + \partial_\nu R_{\lambda \mu \alpha\beta} = 0.
\end{equation}
Contracting $\mu$ and $\alpha$ we obtain
\begin{equation} \label{bianch}
\partial_{\lambda} R_{\nu \beta} + \partial^{\mu} R_{\nu \lambda \mu\beta} - \partial_\nu R_{\lambda\beta}
=0.
\end{equation}
We choose  $\beta$ to be in the  time direction and $\nu, \lambda$ to be space like.
Then the Hamiltonian constraint (\ref{g1}) implies that we obtain  the following constraint that 
must be satisfied by the curvature
\begin{equation} \label{g3}
\partial^\mu R_{0\mu ij} =0.
\end{equation}
Now choose $\nu , \beta$ in (\ref{bianch}) to be in the time direction then the  Hamiltonian constraint
(\ref{g2}) together with the on shell conditions imply
\begin{equation} \label{g4}
\partial^\mu R_{0\mu 0 i} =0.
\end{equation} 

The equations (\ref{g3}) and (\ref{g4}) are analogous to the Gauss law constraint of electrodynamics. 
As we have seen that 
these fields are field strengths of a $U(1)$ gauge potential. 
Therefore the same argument that led to the labelling of superselection sectors 
by the radial component of the electric or magnetic field applies to these curvature components. 
Thus we conclude that superselection sectors are labelled by 
the components $R_{0\hat r ij}, R_{0\hat r 0j}$ of the Riemann tensor.

In the next sub-sections  we will quantize the linearise graviton following the methods of 
\cite{Benedetti:2019uej}. We decompose the graviton in terms of spherical tensor harmonics and fix gauge
and obtain the algebra of gauge-invariant observables. 
We will show that the among the  $6$ components  $R_{0\hat r ij}, R_{0\hat r 0i}$, 
 only the components 
 $R_{0\hat r 0\hat r}$ and $R_{0\hat r \hat e\hat m}$ where $\hat e, \hat m$ are 
angular directions are locally related to the canonical coordinates. 
We then choose these components to label the superselection sectors and 
evaluate their entanglement entropy.  Since both $R_{0\hat r 0\hat r}$ and $R_{0\hat r \hat e\hat m}$
label the superselection sectors, we would need to sum their contribution to the entanglement 
entropy of the edge modes. It is important to note that both these superselection 
sectors arose from the Hamiltonian constraints  (\ref{g1}), (\ref{g2}) of gravity.

\subsection{Lagrangian  in terms of tensor harmonics} \label{laggaug}

In this section, we expand the graviton in terms of tensor harmonics and
decompose the Lagrangian in (\ref{laggrav}) in terms of these modes and fix gauge. 
We follow the methods of  \cite{Benedetti:2019uej} with the difference that we write the tensor harmonics
in terms of covariant tensors rather than cartesian tensors. 
This allows us to apply the methods of tensor calculus.

The spin-2 field is first written as 
\begin{eqnarray} \label{spin2decom}
h_{\mu\nu} =   (h_T)_{\mu\nu} +  (\hat{h}_v)_{\mu\nu} +  ( {h}_s)_{\mu\nu},
\end{eqnarray}
where $h_T$ is the tensor mode which is purely spatial,  $h_v$ is the vector mode which  is non-zero when 
one index is temporal and one spatial, while $h_s$  is non-zero only when both indices are 
temporal. 
The gauge transformation parameter is similarly decomposed as 
\begin{eqnarray}
\xi_\mu = ( \xi_v)_\mu +  ( \xi_s)_\mu.
\end{eqnarray} 
Here $\xi_v$ has only spatial components and $\xi_s$ has only temporal components. 

\subsubsection*{The tensor mode}

 We use tensor harmonics  as a basis of symmetric tensors to expand the tensor mode $h_T$ as 
 follows
\begin{align}
    (h_T)_{\mu\nu}&=\sum_{J,s,\ell,m}h^{J,s}_{(\ell,m)}(T^{J,s}_{\ell,m})_{\mu\nu},
\end{align}
where $T^{J,s}_{(\ell,m)}$ is the tensor harmonics constructed out of the vector harmonics. The covariant form of the tensor harmonics are given by \footnote{ We define  $ [V_\mu \otimes W_\nu]^S  = 
\frac{1}{2} ( V_\mu \otimes W_\nu  + V_\nu  \otimes W_\mu) $  }
\begin{eqnarray} \label{tenhar}
 ( T_{\ell m}^{0l})_{\mu\nu}&=& \hat{r}_{\mu} \otimes Y^{\textbf{r}}_{\ell m,\nu} ,\, \,\,\,\qquad\qquad\qquad\qquad (T_{\ell,m}^{0t})_{\mu\nu}= \frac{Y_{\ell m}(\theta,\phi)}{\sqrt{2}}\left(\delta _{\mu,0} \delta _{0,\nu}-\delta _{\mu,1} \delta _{1,{\nu}}+g_{\mu\nu} \right),\nonumber \\
 (T_{\ell,m}^{1\textbf{e}})_{\mu\nu}&=& \sqrt{2}[\hat{r}_{\mu}\otimes Y^{\textbf{e}}_{\ell,m,\nu}]^S ,\, \,\,\,\qquad\qquad\quad(T_{\ell,m}^{1\mm})_{\mu\nu}= \sqrt{2}[\hat{r}_{\mu}\otimes Y^{\mm}_{\ell,m;\nu}]^S ,\nonumber \\
( T_{\ell,m}^{2\textbf{e}})_{\mu\nu} &=&   \sqrt{\frac{2}{(l-1)(l+2)}} \left[ \left[r \nabla_{\mu} {Y}^{\textbf{e}}_{\ell,m,\nu}  \right]^{S} + \frac{1}{\sqrt{2}} (T^{1\textbf{e}}_{\ell,m})_{\mu\nu}+ \sqrt{\frac{l(l+1)}{2}} (T^{0t}_{\ell,m})_{\mu\nu} \right] ,
\nonumber \\
(T_{\ell m}^{2\mm})_{\mu\nu}&=& \sqrt{\frac{2}{(l-1)(l+2)}}  \left\{ \left[r \nabla_{\mu} Y_{\ell,m,\nu}^{\mm} \right]^{S} + \frac{1}{\sqrt{2}} (T^{1\mm}_{\ell,m})_{\mu\nu} \right\}. 
\end{eqnarray}
The  covariant form of the vector harmonics and the unit orthonormal vectors  are given in  \eqref{defcov} and \eqref{unitvec} respectively. 
Here for the mode  $J =0$ we have $l\geq 0$, $J=1$, $l \geq 1$ and $J=2$, we have  $l \geq 2$. 
Under the gauge transformation the tensor mode of the field transforms as 
\begin{align}
     (h_T)_{\mu\nu}&=\sum_{J,s,\ell,m}h^{J,s}_{(\ell,m)}(T^{J,s}_{\ell,m})_{\mu\nu}+2\sum_{s,\ell,m}\left[\xi^s_{(\ell,m)}\nabla_{\mu}Y^{s}_{\l m ,\nu}+Y^{s}_{\l m ,\mu}\otimes\partial_r\xi^s_{(\ell,m)}\hat{r}_{\nu}\right]^S, \quad s=\r,\e,\m
\end{align}
Here we have also decomposed the vector gauge parameter $\xi_v$   whose index
takes values in the spatial directions in terms of vector harmonics. 
\begin{align}
    (\xi_v)_{\mu}=\sum_{s,\ell,m}\xi^s_{(\ell,m)}Y^{s}_{\l m ,\mu}.
\end{align}
Using the properties of vector and spherical harmonics, one can express the gauge transformed variable $h_T'$ in terms of modes $h^{J,s}_{(\ell,m)}$ and gauge parameters $\xi^s_{(\ell,m)}$.
\begin{align}
  &  (h'_T)_{\mu\nu} = \sum_{\ell,m} \left( h_{(\ell,m)}^{0l}+2\partial_r \xi_{(\ell,m)}^{\r} \right) (T^{0l}_{\ell,m})_{\mu\nu} + \left( h_{(\ell,m)}^{0t}  + \frac{2\sqrt{2}}{r} \xi_{(\ell,m)}^{\r} - \frac{\sqrt{2\ell(\ell+1)}}{r} \xi_{(\ell,m)}^{\e}\right) (T_{\ell,m}^{0t})_{\mu\nu} \nonumber\\
&+\left( h_{(\ell,m)}^{1e} +\frac{\sqrt{2\ell(\ell+1)}}{r}\xi_{(\ell,m)}^{\r} + \sqrt{2}\partial_r \xi_{(\ell,m)}^{\e}  -\frac{\sqrt{2}}{r}\xi_{(\ell,m)}^{\e}\right) T^{1e}_{(\ell,m)}\nonumber\\
 &+ \left( h_{(\ell,m)}^{2e} + \frac{\sqrt{2(\ell-1)(\ell+2)}}{r}\xi_{(\ell,m)}^{\e}\right) T^{2\e}_{(\ell,m)}  + \left( h_{(\ell,m)}^{2\mm} + \frac{\sqrt{2(l-1)(l+2)}}{r}\xi_{(\ell,m)}^{\m} \right) (T^{2\mm}_{\ell,m})_{\mu\nu}\nonumber\\
 & + \left( h_{(\ell,m)}^{1\mm} + \sqrt{2}\partial_r \xi_{(\ell,m)}^{\m} - \frac{\sqrt{2}}{r}\xi_{(\ell,m)}^{\m}  \right) (T^{1\mm}_{\ell,m})_{\mu\nu} .
\end{align}
Note that $\xi_{(\ell,m)}^{\m}$ allows us to cancel the coefficient of $T^{2m}_{(\ell,m)}$, for all $\ell$ and $m$. 
It is clear from thsi tranformation, that we can also use $\xi^{\mm}$  and $\xi^{\r}$ to  eliminate two modes 
out of $h_{(\ell,m)}^{0l}$, $h_{(\ell,m)}^{0t}$ and $h_{(\ell,m)}^{1e}$.
Following \cite{Benedetti:2019uej}, we fix gauge so that one  linear combination of 
these fields remains.  We call this linear combination  the mode 
 $h^{te}_{(\ell,m)}$ 
 Thus fixing gauge we can expand the tensor mode as
\begin{align}\label{tensordecom}
    (h_T)_{\mu\nu} = \sum_{\ell,m} {h}^{0l}_{(\ell,m)} (T^{0l}_{\ell,m})_{\mu\nu}
      + {h}^{te}_{(\ell,m)} \left(\alpha (T^{0t}_{\ell,m})_{\mu\nu} +\beta (T^{1e}_{\ell,m})_{\mu\nu} +\gamma (T^{2e}_{\ell,m})_{\mu\nu} \right)  + {h}^{1\mm}_{\ell,m} (T^{1\mm}_{\ell,m})_{\mu\nu}.
\end{align}
where $\alpha$, $\beta$ and $\gamma$ are some constants. 

\subsubsection*{The vector mode}

The vector mode $(\hat{h} _v)_{\mu \nu}$ is first written as
\begin{equation}
(\hat{h}_v)_{\mu\nu} = [(h_v)_\mu \otimes \hat t_{\nu} ]^S,
\end{equation}
where $\hat t$ is the covariant vector given in (\ref{unitvec}). 
We expand the vector  $h_v$  in terms of vector harmonics given in \eqref{defcov}.
\begin{align}\label{vecdecom}
    (h_v)_{\mu}=\sum_{s\ell m} h_{(\ell,m)}^{0s}(t,r) {Y}_{\ell,m,\mu}^{s}(\theta,\phi).
\end{align}
Let also expand the scalar gauge transformation parameter 
as
\begin{eqnarray}
(\xi_{s})_\mu =\sum_{\ell,m} \xi^0_{(\ell,m)}(t,r) {Y}_{lm}(\theta,\phi)  (\hat t)_\mu.
\end{eqnarray}
Under the gauge transformation, the vector mode transforms as
\begin{align}
    (h'_v)_{\mu}&=\sum_{\ell,m} \left( h_{(\ell,m)}^{0r} + \dot{\xi}^{\r}_{(\ell,m)} + \partial_r \xi^0_{(\ell,m)} \right) Y_{\ell,m,\mu}^{\r} + \left( h_{(\ell,m)}^{0e} + \dot{\xi}^{e}_{(\ell,m)} + \frac{\xi^0_{(\ell,m)}}{r} \right) Y_{\ell,m,\mu}^{\e}\nonumber\\
    &+ \left( h_{(\ell,m)}^{0m} + \dot{\xi}^{\m}_{(\ell,m)}\right) Y_{\ell,m,\mu}^{\m}.
\end{align}
The gauge transformation allows us to choose $\xi^0_{(\ell,m)}$ such that $h'^{0e}_{(\ell,m)}$ vanishes for all $\ell$ and $m$. Using this gauge choice the  vector mode can be written as
\begin{align}
    ({h}_v)_{\mu}=\sum_{\ell m}  {h}_{(\ell,m)}^{0r} Y_{\l,m,\mu}^{\r} + {h}_{(\l,m)}^{0\mm} Y_{\ell,m,\mu}^{\m}.
\end{align}

\subsubsection*{The scalar mode}
Finally the scalar mode is also expanded in  terms of spherical harmonics
\begin{align}\label{scldecom}
    (h_s)_{\mu\nu} &=\sum_{\ell,m}h^{00}_{(\ell,m)}Y_{\ell,m}(\theta,\phi) \hat  t_\mu \otimes \hat t_\nu.
\end{align}

We now substitute the tensor harmonic expansion of 
 $h_{\mu\nu}$ in the Lagrangian given in \eqref{laggrav}. 
 The Lagrangian can be written in two parts \cite{Benedetti:2019uej}. 
 \begin{equation} \label{gravmlag}
    \int d^3 x \mathcal{L} = \sum_{\ell,m =-\l}^\l
    \int_0^\infty dr\left(\mathcal{L}^{(1)}_{\ell,m}+\mathcal{L}^{(2)}_{\ell,m}\right),
\end{equation}
where $\mathcal{L}^{(1)}$ contains only the fields  $h^{1\mm}_{(\ell,m)}$  and the non-dynamical 
modes $h^{0\mm}_{(\ell,m)}$ .  The second part
$\mathcal{L}^{(2)}_{\ell,m} $ involves the fields $h^{0l}_{(\ell,m)}$ and $h^{te}_{(\ell,m)}$ together with the Lagrange multipliers $h^{0\r}_{(\ell,m)}$ and $h^{00}_{(\ell,m)}$.
To write the Lagrangian explicitly in terms of the modes we use the 
following reality property of the tensor harmonics

\begin{align}
    \hat{T}^{J,s\, *}_{\ell,m}=(-1)^m \hat{T}^{J,s}_{\ell,-m}.
\end{align}
Then the reality of the tensor  $h_{\mu\nu}$ leads to the following 
reality conditions obeyed by the modes.
\begin{eqnarray}\label{realitygraviton}
 h^{1\mm}_{(\ell,-m)}&=(-1)^m h^{1\mm *}_{(\ell,m)},\qquad h^{ te *}_{(\ell,m)}&=(-1)^m h^{te *}_{(\ell,m)},\\
 h^{0l}_{(\ell,-m)}&=(-1)^m h^{0l *}_{(\ell,m)},\qquad  h^{0t}_{(\ell,-m)}&=(-1)^m h^{0t *}_{(\ell,m)}.
\end{eqnarray}

\subsubsection*{The Lagrangian ${\cal L}^{(1)}$ and its equations of motion}

The Lagrangian ${\cal L}^{(1)}$ is given by 
\begin{align}\label{Lg1}
     \mathcal{L}^{(1)}_{(\ell,m)}&=\frac{r^2}{2} \dot{h}^{1\mm}_{(\ell,m)}\dot{h}^{1\mm *}_{(\ell,m)} -\frac{(\ell-1)(\ell+2)}{2}h^{1\mm}_{(\ell,m)}h^{1\mm *}_{(\ell,m)} +r^2\partial_r h^{0\mm}_{(\ell,m)}\partial_r h^{0\mm *}_{(\ell,m)} \nonumber \\
&+ \ell(\ell+1)h^{0\mm}_{(\ell,m)}h^{0\mm *}_{(\ell,m)} +\sqrt{2}\dot{h}^{1\mm *}_{(\ell,m)}\left( rh^{0\mm}_{(\ell,m)} - r^2\partial_r h^{0\mm}_{(\ell,m)}\right) .
\end{align}
Here $\l$ runs from $1, 2, \cdots \infty$.  Note that though the cross term between $h^{0\mm}$ and $h^{1\mm}$
appears complex,  the sum over $m$ from $-\l$ to $\l$  in (\ref{gravmlag}) ensures the reality of the 
full Lagrangian. 
The  canonical conjugate momentum to 
 $h^{1\mm}$ by varying the Lagrangian with respect to $\dot{h}^{*1\mm}_{(\ell,m)}$
\begin{align}\label{mom1}
    \pi^{1\mm *}_{(\ell,m)}&=\frac{\partial \mathcal{L}^{(1)}_{(\ell,m)}}{\partial \dot{h}^{1\mm *}_{(\ell,m)}}\nonumber\\
&=    r^2\dot{h}^{1\mm}_{(\ell,m)}+\sqrt{2}(rh^{0\mm}_{(\ell,m)}-r^2\partial_rh^{0\mm}_{(\ell,m)}).
\end{align}

Let us discuss the modes $\l \geq 2$ first. 
From the Lagrangian \eqref{Lg1} we obtain
\begin{align}\label{eom1}
    r^2\ddot{h}^{1\mm}_{(\ell,m)}+\sqrt{2}(r\dot{h}^{0\mm}_{(\ell,m)}-r^2\partial_r\dot{h}^{0\mm}_{(\ell,m)})+(\ell-1)(\ell+2)h^{1\mm}_{(\ell,m)}&=0,
\end{align}
which can also be written as
\begin{align}
      \dot{\pi}^{1\mm *}_{(\ell,m)}+(\ell-1)(\ell+2)h^{1\mm}_{(\ell,m)}&=0.
\end{align}
The mode $h^{0\mm}$ can be eliminated by the constraint equation which is given by
\begin{align}\label{const1}
    \sqrt{2}\left(\partial_r\pi^{1\mm *}_{(\ell,m)}+\frac{\pi^{1\mm *}_{(\ell,m)}}{r}\right)=-2(\ell-1)(\ell+2)h^{0\mm}_{(\ell,m)}.
\end{align}
Finally we get the equation of motion only in $h^{1\mm}_{(\ell,m)}$ variable
\begin{align}
   r^2( \ddot{h}^{1\mm}_{(\ell,m)}- \partial_r^2h^{1\mm}_{(\ell,m)})+\ell(\ell+1)h^{1\mm}_{(\ell,m)}&=0.
\end{align}
We solve this equation of motion by expanding it in Fourier modes in time first and then solve the radial equation. The solution which is regular at the origin is given by 
\begin{align} \label{solh1m}
     h^{1\mm}_{ (\ell,m)}(t,r)=e^{-ikt}a_{1\mm (\ell,m)}(k)\sqrt{r}J_{\ell+\frac{1}{2}}(|k| r).
\end{align}
Here $a_{1\mm (\ell,m)}(k)$ is the integration constant and $J_{\ell+\frac{1}{2}}(|k| r)$ is the Bessel function. The equation of $\pi^{*1\mm}_{(\ell,m)}$ can be obtained from \eqref{mom1}
\begin{align}
     \pi^{1\mm *}_{(\ell,m)}&=
    r^2(-i k e^{-ikt}a_{1\mm (\ell,m)}(k)\sqrt{r}J_{\ell+\frac{1}{2}}(|k| r))+\sqrt{2}(rh^{0\mm}_{(\ell,m)}-r^2\partial_rh^{0\mm}_{(\ell,m)})\nonumber\\
    &= r^2(-i k e^{-ikt}a_{1\mm (\ell,m)}(k)\sqrt{r}J_{\ell+\frac{1}{2}}(|k| r))+\frac{r^2\partial_r^2 \pi^{1\mm *}_{(\ell,m)}-2  \pi^{1\mm *}_{(\ell,m)}}{(\ell-1)(\ell+2)}).
\end{align}
In the second line we have used the constraint equation \eqref{const1} to eliminate $h^{0\mm}_{(\ell,m)}$ from the equation of motion.  Therefore the
equation  for $\pi^{1\mm *}$ becomes an in-homogenous second order equation in the radial coordinate. The general solution is given by
\begin{align}
    \pi^{1\mm *}_{(\ell,m)}(r,t)&=c_1 r^{l+1}+c_2 r^{-l}+(\ell-1)(\ell+2)a_{1\mm (\ell,m)}(k)e^{-ikt}\left(\frac{i r 2^{-\ell -\frac{1}{2}} (k r)^{\ell }}{\sqrt{k} \Gamma \left(\ell +\frac{3}{2}\right)}-\frac{i \sqrt{r}  J_{\ell +\frac{1}{2}}(|k| r)}{k}\right).
\end{align}
Demanding that the solution be regular at the origin yields $c_2=0$ since $\ell\geq 2$ for these
modes. We further demand the regularity of the solution at infinity which fixes $c_1$ and we obtain
\begin{align} \label{solpi1m}
    \pi^{1\mm *}_{(\ell,m)}(r,t)=-\frac{i(\ell-1)(\ell+2)}{k} a_{1\mm (\ell,m)}(k)e^{-ikt} \sqrt{r}J_{\ell+\frac{1}{2}}(|k| r).
\end{align}

For $\l =1$, the equations of motion (\ref{const1}) leads to the following solution 
\begin{equation}
\pi^{1\mm *}_{(1, m)}  = \frac{c}{r},
\end{equation}
where $c$ is a constant. 
Regularity at the origin implies that we have $c=0$ leading to
\begin{equation} \label{lone}
\pi^{1\mm *}_{(1, m)}  = \pi^{1\mm }_{(1, m)} =0.
\end{equation}

\subsubsection*{The Lagrangian ${\cal L}^{(2)}$ and its equations of motion}

We use the reality conditions to express the Lagrangian of the $h^{te}$ mode in terms of field variables
\begin{align}\label{Lg2}
 &   \mathcal{L}^{(2)}_{(\ell,m)} = \frac{r^2}{2} \left(\beta^2-\alpha^2+\gamma^2\right)\dot{h}^{te}_{(\ell,m)}\dot{h}^{te *}_{(\ell,m)}-\sqrt{2}r^2\alpha \dot{h}^{0l}_{(\ell,m)}\dot{h}^{te *}_{(\ell,m)}+ \frac{r^2}{2}\left(\alpha^2-\gamma^2\right)\partial_r h^{te}_{(\ell,m)} \partial_r h^{te *}_{(\ell,m)}\nonumber \\
&+\sqrt{2}\alpha r h^{te}_{(\ell,m)}\partial_r h^{0l *}_{(\ell,m)} +h^{0l}_{(\ell,m)}h^{0l *}_{(\ell,m)}+\left(\beta^2-\frac{\sqrt{\ell(\ell+1)}}{2}\alpha\beta-\frac{\sqrt{(\ell-1)(\ell+2)}}{2}\beta\gamma\right) h^{te}_{(\ell,m)}h^{*te}_{(\ell,m)}\nonumber \\
&+\sqrt{2}\left( \frac{\ell(\ell+1)}{2}\alpha - \sqrt{\ell(\ell+1)}\beta + \frac{\sqrt{(\ell-1)\ell(\ell+1)(\ell+2)}}{2}\gamma \right)h^{0l}_{(\ell,m)}h^{te *}_{(\ell,m)}+\ell(\ell+1)h^{0r}_{(\ell,m)}h^{0r *}_{(\ell,m)} \nonumber \\
&+h^{0r *}_{(\ell,m)}\left[ 4r\dot{h}^{0l}_{(\ell,m)} -2\sqrt{2}\alpha r^2 \partial_r \dot{h}^{te}_{(\ell,m)} - \sqrt{2}\left(2 \alpha + \sqrt{\ell(\ell+1)}\beta\right)r \dot{h}^{te}_{(\ell,m)}\right]+h^{00 *}_{(\ell,m)} \left[-2r\partial_r h^{0l}_{(\ell,m)} \right. \nonumber \\ 
&-\left. (\ell(\ell+1)+2)h^{0l}_{(\ell,m)} +\sqrt{2}\alpha r^2 \partial_r\partial_rh^{te}_{(\ell,m)}+\sqrt{2}\left(3\alpha+\sqrt{\ell(\ell+1)}\beta \right) r \partial_rh^{te}_{(\ell,m)}\right. \nonumber \\
&+ \left. \frac{1}{\sqrt{2}}\left(-(\ell-1)(\ell+2)\alpha+ 4 \sqrt{\ell(\ell+1)}\beta-\sqrt{(\ell-1)\ell(\ell+1)(\ell+2)}\gamma\right)h^{te}_{(\ell,m)} \right].
\end{align}
Here as shown in \cite{Benedetti:2019uej}, $\l =2, 3, \cdots$, 
the Lagrangian for the $\l =1$ mode vanishes. 

Let us use the constraints to simplify the Lagrangian. 
Varying  the action with respect to $h^{00 *}_{(\ell,m)}$ we obtain the constraint
\begin{align}\label{hzeroconstraint}
    &-2r\partial_r h^{0l}_{(\ell,m)}  -(\ell(\ell+1)+2)h^{0l}_{(\ell,m)} +\sqrt{2}\left(3\alpha+\sqrt{\ell(\ell+1)}\beta \right) r \partial_r h^{te}_{(\ell,m)}+\sqrt{2}\alpha r^2 \partial_r\partial_r h^{te}_{(\ell,m)} \nonumber \\
&+ \frac{1}{\sqrt{2}}\left(-(\ell-1)(\ell+2)\alpha+ 4 \sqrt{\ell(\ell+1)}\beta-\sqrt{(\ell-1)\ell(\ell+1)(\ell+2)}\gamma\right)h^{te}_{(\ell,m)} =0 .
\end{align}
The field $h^{0r}_{(\ell,m)}$ is non-dynamical, we can eliminate it using its equations of motion
which is given by 

\begin{align}\label{h0rconstraint}
        \ell(\ell+1)h^{0r}_{(\ell,m)}+\left[ 4r\dot{h}^{0l}_{(\ell,m)} -2\sqrt{2}\alpha r^2 \partial_r \dot{h}^{te}_{(\ell,m)}
      - \sqrt{2}\left(2 \alpha + \sqrt{\ell(\ell+1)}\beta\right)r \dot{h}^{te}_{(\ell,m)}\right]=0.
    \end{align}
    Solving for $h^{0l}$ using (\ref{hzeroconstraint}) will in general result in non-local terms. 
    To obtain a local Lagrangian we follow \cite{Benedetti:2019uej}.. 
   The remaining gauge freedom  allows the ansatz
\begin{align}\label{gaugelocal}
    h^{0l}_{(\ell,m)}=a h^{te}_{(\ell,m)}+br\partial_r h^{te}_{(\ell,m)}.
\end{align}
Substituting  \eqref{gaugelocal} in \eqref{hzeroconstraint}
we obtain

\begin{align}
    &\sqrt{2}\left(\alpha - \sqrt{2} b \right) r^2 \partial_r \partial_r h^{te}_{(\ell,m)} + \left(3\sqrt{2}\alpha + \sqrt{2\ell(\ell+1)}\beta -(\ell(\ell+1)+4)b-2a\right)r\partial_rh^{te}_{(\ell,m)} \nonumber \\
&\frac{1}{\sqrt{2}}\left(-(\ell-1)(\ell+2)\alpha + 4\sqrt{\ell(\ell+1)} \beta - \sqrt{(\ell-1)\ell(\ell+1)(\ell+2)}\gamma \right.\nonumber\\
&\hspace{8cm}\left. -a \sqrt{2}(l(l+1)+2) \right)h^{te}_{(\ell,m)} = 0.
\end{align}
Demanding that 
the coefficients  vanish at each order  in derivatives  in the above equation
we obtain
\begin{align}\label{abc}
    \begin{split}
    a&=\sqrt{\frac{2}{\ell(\ell+1)}} \beta - \sqrt{\frac{(\ell-1)(\ell+2)}{2\ell(\ell+1)}} \gamma,\\
        b&=\sqrt{\frac{2}{\ell(\ell+1)}} \beta + \sqrt{\frac{2}{(\ell-1)\ell(\ell+1)(\ell+2)}} \gamma,\\
        \alpha&=\frac{2}{\sqrt{\ell(\ell+1)}} \beta + \frac{2}{\sqrt{(\ell-1)\ell(\ell+1)(\ell+2)}} \gamma.\\
    \end{split}.
\end{align}
The equation (\ref{h0rconstraint}) and (\ref{gaugelocal}) together with (\ref{abc}) implies that the Lagrangian ${\cal L}^{(2)}$ 
 can be 
written  only in terms of $h^{te}$. This is given by 
\begin{align}
    \mathcal{L}^{(2)}_{\ell,m}&=\frac{\gamma^2}{2}\left(r^2 \dot{h}^{te}_{(\ell,m)}\dot{h}^{te *}_{(\ell,m)}-r^2(\partial_r h^{te}_{(\ell,m)})(\partial_r h^{te *}_{(\ell,m)})-\ell(\ell+1)h^{te}_{(\ell,m)}h^{te *}_{(\ell,m)}\right).
\end{align}
The 
canonical conjugate momentum  to $h^{te}$ is given by 
\begin{align}\label{mom2}
    \pi^{te *}_{(\ell,m)}=\frac{\partial \mathcal{L}^{(2)}}{\partial \dot{h}^{te *}_{(\ell,m)}}=\gamma^2 r^2 \dot{h}^{te}_{(\ell,m)}.
\end{align}
The equation of motion is then given by
\begin{align}
  r^2(\ddot{h}^{te}_{(\ell,m)}-\partial_r^2h^{te}_{(\ell,m)}) -2r\partial_r h^{te}_{(\ell,m)}+\ell(\ell+1)h^{te}_{(\ell,m)}&=0.
\end{align}
We solve this equation of motion by expanding it in Fourier modes in time first and then solve the radial equation. The solution which is regular at the origin is given by 
\begin{align} \label{solhte}
      h^{te}_{(\ell,m)}(t,r)= a_{te (\ell,m)}(k) r^{-\frac{1}{2}}J_{\ell +\frac{1}{2}}(|k| r)e^{-ikt}.
\end{align}
We use  \eqref{mom2} to write the Fourier modes of the  momentum which is given by 
\begin{align} \label{solpite}
   \pi^{te *}_{(\ell,m)}&=-i k\gamma^2 a_{ te (\ell,m)}(k) r^{\frac{3}{2} }J_{\ell +\frac{1}{2}}(|k| r)e^{-ikt} .
\end{align}
Here $a_{te  ( \l, m ) } (k) $ is the arbitrary integration constant for the classical solution.

\subsection{Curvature and the gauge fixed modes} \label{curvcomp}

We have shown that the superselection sectors are in principle 
 labelled by the $6$ curvature tensors of the form $R_{0\hat r ij}, R_{0\hat r 0i}$, where 
 $i, j$ are spatial directions. 
 In this section by explicitly evaluating these curvature components using the gauge discussed in 
 section (\ref{laggaug}), we will see only $2$ are related locally to the canonical coordinates $\pi^{1\mm}$ and 
 $h^{te}$. 
 The gauge choice  is adapted to the spherical symmetry of the problem. 
 It is known that though gauge fixing converts a gauge field to a physical quantity, locality  depends 
 on the gauge choice \cite{Casini:2013rba}. 
To evaluate the curvatures we use Mathematica. It is first convenient to write
the linearized curvature in terms of covariant derivatives. 
This and the writing the tensor harmonics as covariant tensors  (\ref{tenhar}) 
and covariant vectors (\ref{defcov}) allows us to use tensor calculus to 
evaluate the curvature. 
\begin{align} \label{curvten2}
     R_{\mu\nu\rho\sigma}=\frac{1}{2}[\nabla_{\nu}\nabla_{\rho}h_{\mu\sigma}-\nabla_{\mu}\nabla_{\rho}h_{\nu\sigma}+\nabla_{\mu}\nabla_{\sigma}h_{\nu\rho}-\nabla_{\nu}\nabla_{\sigma}h_{\mu\rho}].
\end{align}
We substitute the tensor, vector and scalar mode decomposition given in (\ref{spin2decom})
 \eqref{tensordecom},  \eqref{vecdecom} and \eqref{scldecom} 
 respectively in the expression of the curvature tensor (\ref{curvten2})  and use Mathematica to simplify the 
 calculation \footnote{The Mathematica note book can be found as an ancillary file along with the arXiv version}. 

\subsubsection*{$ R_{\hat{t}\hat{r}\hat{e}\hat{m}}$ }

Consider
$ R_{\hat{t}\hat{r}\hat{e}\hat{m}}$ which is obtained by taking  appropriate projections with contravariant unit vectors
\begin{align}  
     R_{\hat{t}\hat{r}\hat{e}\hat{m}}&=\hat{t}^{\mu}\hat{r}^{\nu}\hat{e}^{\rho}\hat{m}^{\sigma}R_{\mu\nu\rho\sigma}.
     \end{align}
     Substituting the expansions of the metric in terms of its  tensor, vector and scalar modes, we obtain
     \begin{align} \label{magcurv}
      R_{\hat{t}\hat{r}\hat{e}\hat{m}}  &=\sum_{\ell,m =-\l}^\l\frac{\left(2 h^{0\mm}_{(\ell, m)}(t,r)-2 r\partial_r h^{0\mm}_{(\ell, m)}(t,r)+\sqrt{2} r\dot{ h}^{1\mm}_{(\ell, m)}(t,r)\right)}{2 r^2 }\sqrt{\ell(\ell+1)}Y_{\ell m}(\theta,\phi)\nonumber\\
     &=\sum_{\ell m}\sqrt{\frac{\ell(\ell+1)}{2}}\frac{\pi^{1\mm*}_{(\ell,m)}}{r^3}Y_{\ell m}(\theta,\phi).
\end{align}
Here though the sum over $\l$ runs from $\l = 2, \cdots \infty$,  since from (\ref{lone}) we see that
the $\l =1$ component 
$\pi^{1\mm *}_{(1, m ) } $ vanishes.
Note that $ R_{\hat{t}\hat{r}\hat{e}\hat{m}}$ is a gauge-invariant observable and  it is related to the 
 canonical momentum of the mode $h^{1\mm}_{(\ell,m)}$ without any radial derivative. 
 Therefore this relation is local in the radial co-ordinates and the curvature component 
 can be used to label superselection sectors.

\subsection*{ $R_{\hat{t}\hat{r}\hat{t}\hat{r}}$}
To evaluate
 $R_{\hat{t}\hat{r}\hat{t}\hat{r}}$, we  use the vacuum  Einstein equation
\begin{align}
    R_{\hat{r}\hat{r}}=0.
\end{align}
From the equation of motion, we relate the curvature component $R_{\hat{t}\hat{r}\hat{t}\hat{r}}$ to other curvature components which are easy to evaluate
\begin{align}
   R_{\hat{t}\hat{r}\hat{t}\hat{r}}&=g^{\theta\theta}R_{\theta r\theta r}+g^{\phi\phi}R_{\phi r\phi r}\nonumber\\
    &=\sum_{\ell m}\Big[\frac{2 r \partial_rh^{0 l }_{(\ell,m)}(t,r)  +h^{0l}_{(\ell,m)}(t,r) \ell  (\ell +1) }{r^2}
    -\frac{\sqrt{2}\alpha \left(2   \partial_rh^{\text{te}}_{(\ell,m)}(t,r)+  r \partial_r^2h^{\text{te}}_{(\ell,m)}(t,r)\right)}{r}
    \nonumber\\
    &-\beta\sqrt{2\ell(\ell+1)}\frac{\left(h^{te}_{(\ell,m)}(t,r)+\partial_r h^{te}_{(\ell,m)}(t,r)\right)}{r^2}\Big]Y_{\ell,m}(\theta,\phi).
\end{align}
In the last line we substitute the mode expansion of $h_{\mu\nu}$. At this stage, this component of the curvature tensor appears to be  non-local in $r$. 
 But using  the  gauge choice  in \eqref{gaugelocal}  we  relate  $h^{0l}_{(\ell,m)}$ as a function of $h^{te}_{(\ell,m)}$. Finally we replace $a$, $b$ and $\alpha$ in terms of $\beta$ and $\gamma$ given in \eqref{abc} to obtain
\begin{align} \label{eleccurv}
       R_{\hat{t}\hat{r}\hat{t}\hat{r}}= -\sum_{\ell,m}\frac{\gamma  \sqrt{(\ell -1) \ell  (\ell +1) (\ell +2)} h^{te}_{(\ell,m)}(t,r) Y_{\ell,m}(\theta ,\phi )}{\sqrt{2} r^2}.
\end{align}
Here $l$ runs from $\l = 2, 3, \cdots$. 
It is interesting to 
note that, imposing the  gauge condition   \eqref{gaugelocal} 
removes the apparent non-locality of the curvature tensor $  R_{\hat{t}\hat{r}\hat{t}\hat{r}}$. 
This relation also tells us the  canonical coordinate $h^{te}$  is gauge-invariant in our gauge choice. 
Therefore this curvature component can also be used to label super-selection sectors for a spherical entangling 
surface. 

\subsubsection*{$  R_{\hat{t}\hat{r}\hat{t}\hat{m}}$}

Evaluating this curvature component using the same methods we obtain 
\begin{align}
    R_{\hat{t}\hat{r}\hat{t}\hat{m}}&=\sum_{\ell,m}\frac{1}{\sqrt{2(\ell(\ell+1))}}\left(\ddot{h}^{1\mm}_{(\ell,m)}-\sqrt{2}\partial_r\dot{h}^{0\mm}_{(\ell,m)}\right) |Y^{\mm}_{\ell,m}(\theta,\phi) |.
\end{align}
Since it 
 involves  terms containing time derivatives, we can use the 
  we use the equation of motion and the constraint given in \eqref{eom1} and \eqref{const1} to simplify the expression. 
  Note that the curvature component is proportional to the norm of the vector harmonics 
  $Y^{\mm}_{\ell,m}(\theta,\phi) $. 
  We finally obtain
\begin{align}
   R_{\hat{t}\hat{r}\hat{t}\hat{m}}&=-\sum_{\ell,m}\frac{1}{\sqrt{2\ell(\ell+1)}}\left(\frac{1}{r}\partial_r h^{1\mm}_{(\ell,m)}+\frac{(\ell-1)(\ell+2)+1}{r^2}h^{1\mm}_{\ell,m}\right) |Y^{\mm}_{\ell,m}(\theta,\phi)  |.
\end{align}
We observe that the expression of the curvature tensor involves the radial derivative acting on the field. 
Therefore it does not belong to  the local algebra of observables that contribute to the entanglement 
entropy of a spherical entangling surface.

\subsubsection*{ $R_{\hat{t}\hat{r}\hat{r}\hat{m}}$}

This curvature component is given by 
\begin{align}
R_{\hat{t}\hat{r}\hat{r}\hat{m}} 
& =-\sum_{\ell,m}\frac{1}{r^2\sqrt{2\ell(\ell+1)}}\big[\sqrt{2}h^{0\mm}_{(\ell,m)}-\sqrt{2}r\partial_r(r\partial_rh^{0\mm}_{(\ell,m)})+2r\dot{h}^{1\mm}_{(\ell,m)}+r^2\partial_r\dot{h}^{1\mm}_{(\ell,m)}\big]
|Y^{\mm}_{\ell,m}(\theta,\phi)|\nonumber\\
    &=-\sum_{\ell,m}\frac{1}{r^2\sqrt{2\ell(\ell+1)}}\partial_r\pi^{1\mm *}_{(\ell,m)}| Y^{\mm}_{\ell,m}(\theta,\phi)| .
    \end{align}
    To obtain the 
    second line  we have used the 
    definition of the conjugate momentum $\pi^{1\mm}_{(\ell,m)}$ in (\ref{mom1}).
 The radial derivative on the conjugate momentum indicates  that this curvature component also does belong the 
 local algebra of observables in a sphere. 
    
    \subsubsection*{$R_{\hat{t}\hat{r}\hat{r}\hat{e}}$}
    
    We now evaluate the curvature tensor $ R_{\hat{t}\hat{r}\hat{r}\hat{e}}$
     \begin{align}
      R_{\hat{t}\hat{r}\hat{r}\hat{e}}&=  \sum_{\ell,m}\Big[-\frac{(-h^{0r}_{(\ell,m)}+r\partial_rh^{0r}_{(\ell,m)})}{r^2}+\frac{\dot{h}^{0l}_{(\ell,m)}}{r}-\frac{\beta}{r\sqrt{2\ell(\ell+1)}}(2\dot{h}^{te}_{(\ell,m)}+r\partial_r\dot{h}^{te})\Big] |Y^{e}_{\ell,m}(\theta,\phi) |.
    \end{align}
   The expression involves the  non-dynamical field 
   $h^{0r}_{(\ell,m)}$ which can be replaced by the constraint equation given in \eqref{h0rconstraint}. 
    We also use the gauge condition \eqref{gaugelocal} to substitute for $h^{0l}_{(\ell,m)}$ and finally we obtain
    \begin{align}
    & R_{\hat{t}\hat{r}\hat{r}\hat{e}} \nonumber \\
    &=  \left[  - \frac{\gamma  \sqrt{\ell ^2+\ell -2}}{r \sqrt{ 2\l (\ell +1)  }}\partial_rh^{te}_{(\ell,m)}-\Big(\frac{2 \gamma }{\sqrt{ 2 \ell  (\ell +1) \left(\ell ^2+\ell -2\right)}}+\frac{\beta }{\sqrt{ 2\ell  (\ell +1)}}\Big)\partial_r\dot{h}^{te}_{(\ell,m)}\right]|Y^e_{\ell,m}|\nonumber\\
    &= \left[- \frac{\gamma  \sqrt{\ell ^2+\ell -2}}{r \sqrt{ 2 \l (\ell +1) }}\partial_rh^{te}_{(\ell,m)}
    -\Big(\frac{2 \gamma }{\sqrt{ 2 \ell  (\ell +1) (\ell ^2+\ell -2 )}}+\frac{\beta }{ \sqrt{ 2 \ell  (\ell +1)}} \Big) \partial_r\Big(\frac{\pi^{te *}_{(\ell,m)}}{\gamma^2r^2}\Big)\right]|Y^e_{\ell,m}|.
    \end{align}
    Again, the curvature component involves the radial derivatives acting on fields even after using the local gauge condition.  We do have the freedom to set  term containing the derivative of the canonical momentum  $\pi^{te *}$
    to zero be choosing a suitable $\beta$. However there will still remain the term containing the derivative
    of $h^{te}$.  Therefore we conclude that this component of the curvature also does not 
    belong to the local algebra of observables in the sphere and cannot be used to label 
    superselection sectors. 
    
  \subsubsection*{$R_{\hat{t}\hat{r}\hat{t}\hat{e}}$}
    
The last of the 6 curvature components is 
 $R_{\hat{t}\hat{r}\hat{t}\hat{e}}$. We use the vacuum Einstein equation $R_{ri}=0$ to evaluate it, where $i $ denotes the angular coordinates on $S^2$.
    \begin{align}
        R_{\hat{t}\hat{r}\hat{t}\hat{i}}&=g^{ab}R_{\hat{a}\hat{r}\hat{b}\hat{i}}.
    \end{align}
    Therefore we write
    \begin{align}
        R_{\hat{t}\hat{r}\hat{t}\hat{e}}&=\hat{e}^k R_{\hat{t}\hat{r}\hat{t}\hat{k}}\nonumber\\
        &=e^{\theta}g^{\phi\phi}R_{\phi r\phi
\theta}+e^{\phi}g^{\theta\theta}R_{\theta r\theta\phi}.
    \end{align}
    Substituting the expansion of  $h_{\mu\nu}$ in (\ref{tenhar}), we obtain
    \begin{align}
    R_{\hat{t}\hat{r}\hat{t}\hat{e}}&=\Big(\frac{h^{0l}_{(\ell,m)}(t,r)}{r^2}   -\alpha\frac{\partial_r h^{te}_{(\ell,m)}}{\sqrt{2}r}  -\sqrt{\frac{2}{\ell(\ell+1)}}\frac{\beta h^{te}_{(\ell,m)}}{r^2}-\frac{(\ell+1)(\ell-2)}{\sqrt{\ell(\ell+1)}}\gamma \frac{\partial_r h^{te}_{(\ell,m}(t,r)}{2r}\Big)|Y^e_{\ell,m}|.
    \end{align}
    We now we use the relation \eqref{abc} to eliminate $h^{0l}$ 
    \begin{align}
       R_{\hat{t}\hat{r}\hat{t}\hat{e}}&=\Big[-\frac{\gamma  \left(\ell ^2+\ell -2\right)}{2 r \sqrt{\ell  (\ell +1)}}\partial_rh^{te}_{(\ell,m)}(t,r)
       -\gamma \sqrt{\frac{(\ell -1) (\ell +2)}{\ell  (\ell +1)}}\frac{h^{te}_{(\ell,m)}(t,r)}{\sqrt{2}r^2}\Big]|Y^{e}_{\ell,m}|.
    \end{align}
    This is again  contains a radial derivative which cannot be eliminated by further choice of $\beta$ and $\gamma$. 
    Therefore this component does not belong to the algebra of local observables in a sphere and 
    cannot be used to label superselection sectors. 
    
   The explicit evaluation of the curvature components leads us to conclude that our of the $6$ components that 
   satisfy the Gauss law, only $2$ of them,  $  R_{\hat{t}\hat{r}\hat{e}\hat{m}}$ and $ R_{\hat{t}\hat{r}\hat{t}\hat{r}}$
    are related to the algebra of local observables in a sphere. 
   We therefore proceed to evaluate the two-point functions of these  components 
   on the sphere to 
   obtain the contribution of superselection sectors to the entanglement entropy.

\subsection{Quantization of the modes}

We first need to quantize the canonical coordinates $( h^{1\mm}, \pi^{1\mm})$ and $( h^{te}, \pi^{te}) $. 
We have found the solutions to the wave equations of these modes in section (\ref{laggaug}). 
We use these solutions to promote these coordinates to operators and impose canonical commutation 
relations. 

\subsubsection*{The mode $h^{1\mm}_{(\ell,m)}$}

The classical solution obtained  for the Fourier mode  of  $h^{1\mm}_{(\ell,m)}$ and $\pi^{1\mm}_{(\ell,m)}$ 
from the equation of motion 
 in (\ref{solh1m}) and (\ref{solpi1m}) respectively  implies the following mode
 expansion of these fields. 
\begin{align}\label{gravmode1}
\begin{split}
     h^{1\mm}_{(\ell,m)}(t,r)&=\frac{1}{\sqrt{2}}\int_{0}^{\infty} kdk\left(a_{1\mm (\ell,m)}(k)e^{-i k t}+(-1)^m 
     a^{\dagger}_{1\mm (\ell, - m ) }(k)e^{i k t}\right)\sqrt{r} J_{\ell+\frac{1}{2}}(|k| r)\\
     \pi^{1\mm*}_{(\ell,m)}(t,r)&=\frac{(\ell-1)(\ell+2)}{\sqrt{2}}\int_0^{\infty} dk\left(-ia_{1\mm(\ell,m)}(k)e^{-i k t}+i(-1)^m a^{\dagger}_{ 1\mm (\ell, -m)}(k) e^{ikt}\right) \sqrt{r} J_{\ell+\frac{1}{2}}(|k| r) .
\end{split}
\end{align}
Here $\l\geq 2$, 
the mode expansion obeys the reality condition given in \eqref{realitygraviton}. Let us also give the mode expansion of $\pi^{1\mm}_{(\ell,m)}$
\begin{align}
\pi^{1\mm}_{\ell m}(t,r)&=\frac{(\ell-1)(\ell+2)}{\sqrt{2}}\int_0^{\infty} dk\left(ia^{\dagger}_{1\mm (\ell,m)}(k)e^{i k t}-i(-1)^m a_{1 \mm(\ell, - m)}e^{- ikt}\right)\sqrt{r} J_{\ell+\frac{1}{2}}(|k| r).
\end{align}
We now promote the variables $h^{1\mm}_{(\ell,m)}$ and $\pi^{1\mm}_{(\ell,m)}$ to the operators which implies $a^{1\mm}_{(\ell,m)}$ and $a^{\dagger1\mm}_{(\ell,m)}$ are also operators. The equal time commutation relation of the conjugate operators are given by
\begin{align}\label{eqtimecomgrav}
    [h^{1\mm}_{(\ell,m)}(t,r),\pi^{1\mm}_{(\ell',m')}(t,r')]=i\delta(r-r')\delta_{\ell \ell'}\delta_{m,m'}.
\end{align}
From the mode expansions given in \eqref{gravmode1} one can see the equal time commutation relation yields the commutation relation of  creation and  annihilation operators.
\begin{align}\label{creationannhilationgrav1}
     [a_{1\mm (\ell,m) }(k),a^{\dagger}_{1\mm (\ell',m') }(k')]=\frac{\delta(k-k')\delta_{\ell\ell'}\delta_{m,m'}}{(\ell+2)(\ell-1)} .
\end{align}
All other commutation relations are trivial. To show the commutation relation of  creation and  annihilation operator we substitute the mode expansion \eqref{gravmode1} in \eqref{eqtimecomgrav} and use the closure relation of the Bessel function given in \eqref{Besselclosure}.

\subsubsection*{The mode $h^{te}_{(\ell,m)}$}
From the classical solution of the equation of motion (\ref{solhte}) and (\ref{solpite}), 
we write the mode expansion of $ h^{te}_{(\ell,m)}$ and $ \pi^{te}_{(\ell,m)}$
\begin{align}
    h^{te}_{(\ell,m)}(t,r)&=\frac{r^{-\frac{1}{2}}}{\sqrt{2}}\int_0^{\infty} dk\left(a_{ te (\ell,m)}(k)e^{-i k t}+(-1)^m
    a^{\dagger}_{ te (\ell, - m) }(k)e^{ikt}\right)  J_{\ell+\frac{1}{2}}(|k|r) \\ \nonumber
    \pi^{te*}_{(\ell,m)}(t,r)&=\frac{\gamma^2 r^\frac{3}{2} }{\sqrt{2}} \int_0^{\infty}k dk\left(-ia_{te (\ell,m) }(k)e^{-i k t}+(-1)^mia^{\dagger }_{ te (\ell, -m) }(k)e^{i k t}\right) J_{\ell+\frac{1}{2}}(|k|r) .
\end{align}
We have used $\pi^{*te}_{(\ell,m)}=\gamma^2 r^2\dot{h}^{te}_{(\ell,m)}$ to write the momentum mode expansion.
The mode expansion of $\pi^{te}_{(\ell,m)}$
\begin{align}
    \pi^{te}_{(\ell,m)}=\frac{\gamma^2 r^{\frac{3}{2}} }{\sqrt{2}} \int_0^{\infty}k dk\left(ia^\dagger_{te (\ell,m) }(k)e^{i k t}-(-1)^mi a_{te (\ell, -m) }(k)e^{- i k t}\right) J_{\ell+\frac{1}{2}}(|k|r) .
\end{align}
Now one promotes the variables  $ h^{te}_{(\ell,m)}$ $\pi^{te}_{(\ell,m)}$ to operators and imposes the equal time commutation relation.
\begin{align}
    [h^{te}_{(\ell,m)}(t,r),\pi^{te}_{\ell}(t,r')]
    &=i\delta(r-r')\delta_{\ell\ell'}\delta_{mm'}.
\end{align}
This implies the commutation relation of creation and  annihilation operators.
\begin{align}\label{creationannhilationgrav2}
     [a_{ te ( \ell m) }(k),a^{\dagger }_{ te (\ell',m') }(k')]=\frac{\delta(k-k')\delta_{\ell\ell'}\delta_{m,m'}}{\gamma^2} .
\end{align}

\subsection{Entanglement entropy of the edge states}

We compute the two-point function of the normal components of the curvature tensors which label the super-selection sector.  As we have seen in section (\ref{curvcomp}), only the curvature components 
 $  R_{\hat{t}\hat{r}\hat{e}\hat{m}}$ and $ R_{\hat{t}\hat{r}\hat{t}\hat{r}}$ are related locally to the 
algebra of gauge-invariant operators in a sphere.

\subsubsection*{Two-point function of $R_{\hat{t}\hat{r}\hat{e}\hat{m}}$  }

From (\ref{magcurv}) we see that 
that $R_{\hat{t}\hat{r}\hat{e}\hat{m}}$ is related to the momentum field $\pi^{1\mm *}$
\begin{align}
 R_{\hat{t}\hat{r}\hat{e}\hat{m}}    
 &=\sum_{\ell, m=-\l}^{ \l}\sqrt{\frac{\ell(\ell+1)}{2}}\frac{\pi^{1\mm *}_{(\ell,m)}}{r^3}Y_{\ell m}(\theta,\phi).
\end{align}
The sum over $\l$ runs from $\l=2, 3, \cdots$. 
Substituting the 
 mode expansion of $\pi^{1\mm*}$ from (\ref{gravmode1}) and using the canonical computation relations (\ref{creationannhilationgrav1}) 
 we compute the two-point function of $ R_{\hat{t}\hat{r}\hat{e}\hat{m}}$
\begin{eqnarray}
&& \langle 0| R_{\hat{t}\hat{r}\hat{e}\hat{m}} (t,  r,  \theta, \phi)   R_{\hat{t'}\hat{r'}\hat{e'}\hat{m'}} ( t, r',  \theta', \phi ') |0\rangle  = 
\\ \nonumber
&& \frac{1}{ 2 ( rr')^{\frac{5}{2} } }  \sum_{\l, \l', m, m'}  (\l+2)(\l-1)(\l'+2)(\l'-1)\left(\l(\l+1)\l'(\l'+1)\right)^{\frac{1}{2}} \int_0^\infty dk dk' 
\left[ 
J_{\l + \frac{1}{2} } ( k r)  J_{\l' + \frac{1}{2} } ( k' r')    \right. \\ \nonumber
& & \qquad\qquad\qquad\qquad \left. 
 \times  (-1)^{m'} 
 \langle 0| a_{ 1\mm (\l, m ) }  (k) a^{\dagger}_{  1\mm  ( \l', m' ) }(k') |0\rangle Y_{\l, m } ( \theta, \phi) Y_{\l', - m'} (\theta', \phi')
 \right].
\end{eqnarray}
Using the commutation relations in (\ref{creationannhilationgrav1}), we obtain
\begin{eqnarray}
 && \langle 0|R_{\hat{t}\hat{r}\hat{e}\hat{m}} (t,  r,  \theta, \phi)   R_{\hat{t'}\hat{r'}\hat{e'}\hat{m'}} ( t, r',  \theta', \phi ') |0\rangle  = \\ \nonumber
 && \qquad\qquad
 \frac{1}{ 2 ( rr')^{\frac{5}{2}} }\sum_{\l, m }(\ell+2)(\ell-1)\ell(\ell+1)  \int_0^\infty  dk  J_{\l + \frac{1}{2} } ( k r)  J_{\l' + \frac{1}{2} } ( k' r')  
 Y_{\l, m } ( \theta, \phi) Y_{\l,  m}^* (\theta', \phi').
\end{eqnarray}
We now use the identity \eqref{besselidentity} to perform the integral over Bessel function and obtain
\begin{eqnarray} \label{maggravcor}
 && \langle 0|R_{\hat{t}\hat{r}\hat{e}\hat{m}} (t,  r,  \theta, \phi)   R_{\hat{t'}\hat{r'}\hat{e'}\hat{m'}} ( t, r',  \theta', \phi ') |0\rangle  = \\ \nonumber
 && \qquad\qquad \qquad\qquad \frac{1}{2 \pi r^3r'^3}
  \sum_{\l m }( \ell+2)(\ell-1) \ell(\ell+1)  Q_\l(\frac{r^2+r'^2}{2rr'})    Y_{\ell m}(\theta,\phi) Y^{*}_{\ell m}(\theta',\phi') .
\end{eqnarray}
We need the two-point functions at the same radial point. In this limit, the expansion of the Legendre function of the second kind is given in \eqref{Qexpand}. Substituting that in the expression of the two-point function
and keeping only the leading term we obtain
\begin{equation} \label{coingravmag}
\lim_{\delta\rightarrow 0} G_{rr} ( r, r+\delta; x, y) =  \frac{1}{4\pi r^6} \log ( \frac{r^2}{\delta^2} ) 
 \sum_{\l\geq 2 , m }   \l(\l +1)(\l-1)(\l+2) Y_{\l, m } ( \theta, \phi) Y_{\l, m }^*( \theta'\, \phi').
\end{equation}

\subsection*{ Two-point function of $ R_{\hat{t}\hat{r}\hat{t}\hat{r}}$  }

The second curvature component locally related to the canonical coordinate is 
 $R_{\hat{t}\hat{r}\hat{t}\hat{r}}$ which is given by (\ref{eleccurv})
\begin{align}
      R_{\hat{t}\hat{r}\hat{t}\hat{r}}= -\sum_{\ell,m =-\l}^\l
      \frac{\gamma  \sqrt{(\ell -1) \ell  (\ell +1) (\ell +2)} h^{te}_{(\ell,m)}(t,r) Y_{\ell,m}(\theta ,\phi )}{\sqrt{2} r^2}.
\end{align}
The sum over $\l$ again runs from $\l=2, 3\cdots$. 
Using the mode expansion of $h^{te}$  in (\ref{eleccor}), 
we compute the two-point function of $  R_{\hat{t}\hat{r}\hat{t}\hat{r}}$
\begin{eqnarray}
&& \langle 0|  R_{\hat{t}\hat{r}\hat{t}\hat{r}} (t,  r,  \theta, \phi)   R_{\hat{t}\hat{r}\hat{t}\hat{r}}( t, r',  \theta', \phi ') |0\rangle  = 
\\ \nonumber
&& \frac{\gamma^2}{ 2 ( rr')^{\frac{5}{2} } }  \sum_{\l, \l', m, m'}  \left(\l(\l+1)(\l-1)(\l+2)\l'(\l'+1)(\l'-1)(\l'+2)\right)^{\frac{1}{2}}  \int_0^\infty dk dk' 
\left[ 
J_{\l + \frac{1}{2} } ( k r)  J_{\l' + \frac{1}{2} } ( k' r')    \right. \\ \nonumber
& & \qquad\qquad\qquad\qquad \left. 
 \times  (-1)^{m'} 
 \langle 0| a_{ te (\l, m ) }  (k) a^{\dagger}_{ te ( \l', m' ) }(k') |0\rangle Y_{\l, m } ( \theta, \phi) Y_{\l', - m'} (\theta', \phi')
 \right].
\end{eqnarray}
Using the commutation relations in (\ref{creationannhilationgrav2}), we obtain
\begin{eqnarray}
 && \langle 0| R_{\hat{t}\hat{r}\hat{t}\hat{r}}(t,  r,  \theta, \phi)   R_{\hat{t'}\hat{r'}\hat{e'}\hat{m'}} ( t, r',  \theta', \phi ') |0\rangle  = \\ \nonumber
 && \qquad\qquad
 \frac{1}{ 2 ( rr')^{\frac{5}{2}} }\sum_{\l, m }(\ell+2)(\ell-1)\ell(\ell+1)  \int_0^\infty  dk  J_{\l + \frac{1}{2} } ( k r)  J_{\l' + \frac{1}{2} } ( k' r')  
 Y_{\l, m } ( \theta, \phi) Y_{\l,  m}^* (\theta', \phi'). 
\end{eqnarray}
We now use the identity \eqref{besselidentity} to perform the integral over Bessel function and obtain
\begin{eqnarray} \label{elecgravcor}
 && \langle 0| R_{\hat{t}\hat{r}\hat{t}\hat{r}} (t,  r,  \theta, \phi)   R_{\hat{t'}\hat{r'}\hat{e'}\hat{m'}} ( t, r',  \theta', \phi ') |0\rangle  = \\ \nonumber
 && \qquad\qquad \qquad \frac{1}{2\pi (rr')^3}
 \sum_{\l, m}
 (\ell+2)(\ell-1)\ell(\ell+1) Q_{\ell}  (\frac{r^2+r'^2}{2rr'})    Y_{\ell m}(\theta,\phi) Y^{*}_{\ell m}(\theta',\phi'). 
\end{eqnarray}
We need the two-point functions at the same radial point. In this limit, the expansion of the Legendre function of the second kind is given in \eqref{Qexpand}. Substituting that in the expression of the two-point function, we obtain
the leading contribution to be given by 
\begin{equation} \label{coingrav2}
\lim_{\delta\rightarrow 0} G_{rr} ( r, r+\delta; x, y) = \frac{1}{4\pi r^6}   \log ( \frac{r^2}{\delta^2} ) 
\sum_{\l \geq 2, m } 
\l(\l +1)(\l-1)(\l+2)  Y_{\l, m } ( \theta, \phi) Y_{\l, m }^*( \theta'\, \phi').
\end{equation}

From (\ref{maggravcor}) and (\ref{elecgravcor} )we see that  the correlators 
 correlators coincide. 
This is expected since the theory of linearised graviton, satisfying vacuum Einstein equations 
are invariant under the `electric-magnetic'  duality  \cite{Casini:2003kf,Casini:2021vmi}
\begin{equation}
\tilde R_{\mu\nu\rho\sigma} = \frac{1}{2} \epsilon_{\rho\sigma \alpha\beta } R_{\mu\nu}^{\;\;\;\alpha\beta}.
\end{equation}

\subsection*{Entanglement of the superselection sectors}

From the discussion in section (\ref{unived}), to determine the logarithmic coefficient of the 
entanglement entropy of the superselection sectors we  see that we need to evaluate the log-determinant
of the leading contribution of the radial coincident Green's function. 
Therefore, 
from (\ref{unived}) and (\ref{coingravmag}) or (\ref{coingrav2}),  the entanglement entropy of 
the superselections sectors determined by the curvature components $R_{\hat{t}\hat{r}\hat{e}\hat{m}}$ 
or  $ R_{\hat{t}\hat{r}\hat{t}\hat{r}}$  
 is given by 
   \begin{equation}
S_{\rm edge} ( \rho_A ) =  \frac{1}{2} \sum_{\l=2}^\infty ( 2\l +1) \log\big[\l ( \l +1)(\l-1)(\l+2) \big].
\end{equation} 
Note that the correlators  in  (\ref{coingravmag}) or (\ref{coingrav2} are diagonal in scalar spherical 
harmonic basis. The diagonal elements 
are independent of the quantum number $m$, this implies that  each eigen value contributes with a 
multiplicity of $(2\l +1)$ to the log-determinant. 

To evaluate the coefficient of the  logarithmic divergence we open up the logarithm 
and write each term using the identity 
\footnote{This method of studying the sums involved in the log-determinant 
can also be used to obtain the logarithmic
contribution of the $U(1)$ theory.}
\begin{equation}
-\log (y) = \int_0^\infty \frac{dt}{t} ( e^{-y t} - e^{-t} ) .
\end{equation}
Using this representation of the logarithm, we obtain 
\begin{equation} \label{edgegrav1}
S_{\rm edge} ( \rho_A ) = -  \int_0^\infty \frac{dt}{2t}  \sum_{\l =2}^\infty
( 2\l +1) \Big( e^{-( l-1)t } +  e^{-\l t}  + e^{ - ( \l +1)t }  +
e^{ - ( \l +2) t  } -4  e^{-t} \Big). 
\end{equation}
The last term involves the sum  over degeneracies
\begin{equation}
g= \sum_{\l =2}^\infty  ( 2\l + 1) .
\end{equation}
To perform this sum 
we resort  to `dimensional regularization' which was introduced in \cite{Giombi:2015haa}, for more details see
around equation 2.7 of \cite{David:2021wrw}. 
Here one performs the sum of degeneracies of scalar harmonics in sufficiently negative  dimensions
for which the sum is convergent  and then 
continues  the result analytically  to positive dimensions. 
This results in 
\begin{equation}
\sum_{\l =2}^\infty ( 2l +1) = -4.
\end{equation}
Substituting this result and performing the  rest of the sums in (\ref{edgegrav1}), we obtain
\begin{equation}
S_{\rm edge} ( \rho_A ) = -  \int_0^\infty \frac{dt}{2t}  \left( 
\frac{ e^{ -t} ( 1+ e^{-t} )( 1 + e^{-2 t} ) ( 5 - 3 e^{-t}) }{ (1- e^{-t} )^2} +16  \right) .
\end{equation}
One can easily  extract the coefficient of the logarithmic divergence from this representation by examining the 
coefficient of the $1/t$ term in the integrand. 
We obtain  \footnote{We have also performed this computation using the methods of \cite{Anninos:2020hfj} 
and obtained the 
same result.}
\begin{equation}
S_{\rm edge} ( \rho_A )|_{\rm log\;coefficient} =  -\frac{8}{3} .
\end{equation}

As we have discussed earlier the Gauss of gravity or the Hamiltonian constraints  (\ref{g1}), (\ref{g2}) results in 
the  equations (\ref{g3}), (\ref{g4}) which determine the superselections sectors. 
Note that these constraints are on the same footing as the physical state condition 
(\ref{u1glaw}) in the $U(1)$ theory. 
We need to sum the contributions arising from all possible superselection sectors 
arising out of the conditions (\ref{g3}), (\ref{g4}). 
We have demonstrated that only $2$ of these curvatures can be written locally in the gauge invariant observables 
on the sphere. Therefore we need to sum the contributions to the entanglement entropy from 
the Gauss law which constrains the curvature components  $  R_{\hat{t}\hat{r}\hat{e}\hat{m}}$ and $ R_{\hat{t}\hat{r}\hat{t}\hat{r}}$. 
This leads to the conclusion that the logarithmic coefficient of the 
gravitational edge modes for a spherical entangling surface is given by 
\begin{equation}
S_{\rm gravitational\,edge modes }(\rho_A)| = -\frac{16}{3} \log \frac{R}{\epsilon}.
\end{equation}
Again we emphasize that this contribution resulted from our choice of the centre
in which the superselection sectors are labelled  by 
the curvature components $  R_{\hat{t}\hat{r}\hat{e}\hat{m}}$ and $ R_{\hat{t}\hat{r}\hat{t}\hat{r}}$.
It will be interesting to study the linearised graviton theory in more detail, so that we can 
prove a trivial  centre can be chosen  just as in the case of the $U(1)$ theory \cite{Casini:2013rba}.

We compare the 
 coefficient of the edge modes of the linearized graviton to the edge partition function 
of the massless spin-2 theory on $S^4$. 
This was evaluated in \cite{Anninos:2020hfj} \footnote{ These can be read out from equations 5.8 and 5.11 of
\cite{Anninos:2020hfj}. 
Note that the last term in 5.11 does not contribute to the logarithmic coefficient. }
and is given by the following integral over the Harish-Chandra character
\begin{eqnarray}
\log {\cal Z}_2 [S^4] &=& \int_0^\infty \frac{dt}{2t} \frac{ 1+q}{ 1-q} \left( [\hat{\chi}_{{\rm bulk}, 2}]_+ - 
[\hat{\chi}_{{\rm edge}, 2}]_+ \right) , \\ \nonumber
[\hat{\chi}_{{\rm bulk}, 2}]_+ &=& \frac{ 10 q^3 -6 q^4}{ (1- q)^3} ,  \qquad \qquad q = e^{-t},  \\ \nonumber
[\hat{\chi}_{{\rm edge}, 2}]_+ &=& \frac{ 10 q^2 - 2 q^3 }{ ( 1- q) }.
\end{eqnarray}
Extracting the contribution of the edge character to the logarithmic coefficient we find
\begin{equation}
\left. \log {\cal Z}_2 [S^4] \right|_{\rm log\; coefficient, \; edge}  = -\frac{16}{3} .
\end{equation}

Just as in the  $U(1)$ case, for the graviton, the logarithmic coefficient of 
edge partition function of the graviton on the sphere agrees with that of the superselection 
sectors of the graviton.   
It will be interesting to understand the relationship between the contribution of superselection to the 
entanglement entropy to that of the edge partition function on spheres further. 
One direction would be to extend the methods in this paper to higher spin fields. 
The coefficient from the edge partition function obtained from the Harish-Chandra character of 
a massless spin-$s$ field in $d=4$ dimension is  given by $-\frac{s^4}{3}$ \cite{David:2020mls}. 

\section{Conclusions} \label{conclusion}

The extractable entanglement or the bulk entanglement of the linearised graviton across a spherical 
surface 
was first evaluated in \cite{Benedetti:2019uej}. 
Decomposing the spin-2 field into tensor harmonics it was shown that the algebra of 
gauge-invariant operators is equivalent to two scalars fields with their $\l=0$ and 
$\l=1$ modes removed. The  logarithmic coefficient 
is given by $-\frac{61}{45}$.  Furthermore, 
considering the spin-2 field on hyperbolic cylinders  and evaluating 
the entanglement entropy also reproduces this coefficient  \cite{David:2020mls}
\footnote{Indeed, we have verified  that even the R\'{e}nyi entropies of two scalar fields with 
their $\l=0$ and 
$\l=1$ modes removed also precisely coincides with that evaluated from considering the spin-2 field 
on the hyperbolic cylinder.}.
However an evaluation of the contribution of the superselection sectors  to the entanglement entropy 
of the linearized 
graviton 
was missing in the literature. 
In this paper we have used  the method of decomposing 
the spin-2 fields into tensor harmonics and fixing a gauge which 
respects the spherical symmetry of the problem developed in 
\cite{Benedetti:2019uej} to evaluate the  logarithmic coefficient of the edge modes or the 
superselection sectors resulting from the Gauss law of gravity. 
One crucial ingredient in the calculation was to determine which among the 
curvature components satisfying the Gauss law of gravity were locally related to the 
algebra of gauge-invariant operators in the sphere.   

 Our choice of superselection sectors resulting 
from the Gauss law picks out a centre for the algebra of local operators for the graviton. 
It would be interesting to develop the extended Hilbert space definition of 
entanglement entropy defined for gauge theories in  \cite{Buividovich:2008gq,Donnelly:2011hn,Donnelly:2014gva,Ghosh:2015iwa,Aoki:2015bsa}
 in detail for the linearized graviton to determine which choice of centre results from the 
extended Hilbert space definition. 
 Here the general methods developed  in  gravity to define 
subregions by  \cite{Donnelly:2016auv,Speranza:2017gxd,Camps:2018wjf}  
 would be useful.   

The methods developed in this paper can be extended to  other fields and to 
 other dimensions. In particular it would be interesting to 
study the contribution of the edge modes of   $p$-forms  in 
arbitrary even dimensions using this approach and verify if their contribution
to the logarithmic coefficient agrees with that from the edge  Harish-Chandra character 
of the sphere partition function. 
As we have seen in this paper this agreement is true for the $U(1)$ fields in arbitrary even dimensions.
Similar questions can be addressed for higher spin fields as well. 

Finally, the entanglement entropy of non-abelian gauge fields contains an additional edge term.
This additional term is tied to the fact that irreducible representations of the superselection 
sectors in the non-Abelian theories
have dimensions greater than unity \cite{Donnelly:2014gva,Soni:2015yga}.  
Recently, the authors of 
\cite{Takayanagi:2019tvn} studied the Hayward term in gravity and suggested that the Hayward term corresponds to the  edge entanglement 
 associated with the above additional contribution in the graviton theory. 
 This occurs in the full non-linear theory. 
It would be interesting to study this further using the methods of  \cite{Donnelly:2016auv,Speranza:2017gxd,Camps:2018wjf}.

\appendix 
\section{Electric correlator in the Coulomb gauge }  \label{appen}

The two-point functions of the electric field on sphere  gauge-invariant and therefore it should not depend on 
gauge choices. 
 In \cite{Soni:2016ogt}, the two-point function of the electric field was evaluated in the  Coulomb gauge ,i.e, $A_0=0$ and $\nabla. \vec{A}=0$. In this paper we have chosen a different gauge which is $A^{'\e}_{(\ell,m)}=0$ for all $\ell$ and $m$. Here $A^{\e}$ is the component of the vector field obtained by taking the projection
\begin{align}
    A^{\e}_{(\ell,m)}=e^{\mu}A_{\mu,(\ell,m)}.
\end{align}
In this appendix we would like to compare the two-point obtained in 
equation A.12 of \cite{Soni:2016ogt} with what 
 we obtain in \eqref{eleccor}. 
 Equation A.12 
  \cite{Soni:2016ogt}  reads
\begin{align}\label{corsoni}
    G_{rr'}&=\frac{1}{\pi^2(r^2+r'^2)^2}\sum_{\ell=0}^{\infty}(4\ell+1)\Big\{\sum_{n=0}^{\infty}\left(\frac{1-\alpha^2}{\alpha}n-\alpha\right)\frac{2n!!}{(2n-2\ell)!!}\frac{(2n+1)!!}{(2n+2\ell+1)!!}\alpha^{2n}\Big\}P_{2\ell}\nonumber\\
    &+\sum_{\ell=0}^{\infty}(4\ell+3)\Big\{\sum_{n=0}^{\infty}\left(\frac{1-\alpha^2}{\alpha}n-\frac{3\alpha^2-1}{2\alpha}\right)\frac{(2n+2)!!}{(2n-2\ell)!!}\frac{(2n+1)!!}{(2n+2\ell+3)!!}\alpha^{2n+1}\Big\}P_{2\ell+1}.
\end{align}
Here $P_{\ell}$ is the Legendre polynomial related to the spherical harmonics by the following relation
\begin{align}
    P_{\ell}(\cos \gamma)&=\sum_{m=-\l}^{\l}\frac{4\pi}{2\ell+1}Y_{\ell m}(\theta, \phi )Y^{*}_{\ell m}(\theta'\phi').
\end{align}
where $\gamma$ is the angle between the unit vectors determined by the angular coordinates on the sphere.
$\alpha$ is related to the radial coordinates $r$ and $r'$ by the following relation
\begin{align}
   \frac{2rr'}{r^2+(r')^2}=\alpha.
\end{align}
Note that equation (\ref{corsoni}) is a series in $\alpha$ for every $\l$, while the expression 
 in (\ref{eleccor})  is a closer function of $\alpha$. 

To compare the two correlators we expand
 \eqref{eleccor} around $\alpha=0$ and compare it with \eqref{corsoni} for each $\ell$ values.
Expanding the two pint function given in \eqref{eleccor} around $\alpha=0$, we obtain
\begin{align}\label{corexpand}
&\langle 0| F_{\hat t \hat r} (t,  r,  \theta, \phi)   F_{\hat t' \hat r'} ( t, r',  \theta', \phi ') |0\rangle  
 =\nonumber\\
&\sum_{\ell,m}\frac{Y_{\ell m}(\theta, \phi)Y_{\ell m}^{*}(\theta', \phi ')\alpha^{\ell}}{\sqrt{\pi}(rr')^2}\Big[\frac{\alpha  2^{-\ell -2} \ell  (\ell +1) \Gamma (\ell +1)}{ \Gamma \left(\ell +\frac{3}{2}\right)}+\frac{\alpha ^3 2^{-\ell -3} \ell  (\ell +1)^2 (\ell +2) \Gamma (\ell +1)}{  (2 \ell +3)  \Gamma \left(\ell +\frac{3}{2}\right)}\nonumber\\
&+\frac{\alpha ^5 2^{-\ell -5} \ell  (\ell +1)^2 (\ell +2) (\ell +3) (\ell +4) \Gamma (\ell +1)}{  (2 \ell +3) (2 \ell +5) \Gamma \left(\ell +\frac{3}{2}\right)}\nonumber\\
&+\frac{\alpha ^7 2^{-\ell -6} \ell  (\ell +1)^2 (\ell +2) (\ell +3) (\ell +4) (\ell +5) (\ell +6) \Gamma (\ell +1)}{3 (2 \ell +3) (2 \ell +5) (2 \ell +7) \Gamma \left(\ell +\frac{3}{2}\right)}+\cdots\Big].
\end{align}
Now  we are in a position to compare the series in  $\alpha$ for each value of $\ell$ with \eqref{corsoni}.

\paragraph{$\ell=0$}, Both correlators vanish at $\ell=0$.
\paragraph{$\ell=1$}, we substitute $\ell=1$ in \eqref{corexpand} and obtain
\begin{align}
\langle 0| F_{\hat t \hat r} (t,  r,  \theta, \phi)   F_{\hat t' \hat r'} ( t, r',  \theta', \phi ') |0\rangle _{|\ell=1}&=  
\frac{Y_{1 m}(\theta, \phi)Y_{1  m}^{*}(\theta', \phi ')}{\pi ( rr')^2}
\left( 
  \frac{\alpha ^2}{3 }+\frac{\alpha ^4}{5 }+\frac{\alpha ^6}{7 }+\frac{\alpha ^8}{9 }+ \cdots 
  \right) .
\end{align}
From \eqref{corsoni} we obtain
\begin{align}
    G_{rr'}|_{\ell=1}=
    \frac{Y_{1 m}(\theta, \phi)Y_{1 m}^{*}(\theta', \phi ')}{\pi ( rr')^2}
\left( 
  \frac{\alpha ^2}{3 }+\frac{\alpha ^4}{5 }+\frac{\alpha ^6}{7 }+\frac{\alpha ^8}{9  }+ \cdots 
  \right) .
\end{align}

Similarly we have checked the expansion  for $\ell=2$ to $\ell=5$ 
\begin{align}
\langle 0| F_{\hat t \hat r} (t,  r,  \theta, \phi)   F_{\hat t' \hat r'} ( t, r',  \theta', \phi ') |0\rangle _{|\ell=2} &=   
   \frac{Y_{2 m}(\theta, \phi)Y_{2  m}^{*}(\theta', \phi ')}{\pi ( rr')^2} \left( 
 \frac{2 \alpha ^3}{5}+\frac{12 \alpha ^5}{35}+\frac{2 \alpha ^7}{7 }+\frac{8 \alpha ^9}{33 }+\cdots \right) 
 \nonumber\\
&= G_{rr'}|_{\ell=2}.
\end{align}

\begin{align}
\langle 0| F_{\hat t \hat r} (t,  r,  \theta, \phi)   F_{\hat t' \hat r'} ( t, r',  \theta', \phi ') |0\rangle _{|\ell=3}&=
  \frac{Y_{3  m}(\theta, \phi)Y_{3  m}^{*}(\theta', \phi ')}{\pi ( rr')^2} \left( 
\frac{12 \alpha ^4}{35}+\frac{8 \alpha ^6}{21}+\frac{4 \alpha ^8}{11 }+\frac{48 \alpha ^{10}}{143 }+\cdots
\right)\nonumber\\
&= G_{rr'}|_{\ell=3}.
\end{align}

\begin{align}
\langle 0| F_{\hat t \hat r} (t,  r,  \theta, \phi)   F_{\hat t' \hat r'} ( t, r',  \theta', \phi ') |0\rangle _{|\ell=4}&=    
  \frac{Y_{4  m}(\theta, \phi)Y_{4  m}^{*}(\theta', \phi ')}{\pi ( rr')^2} \left( 
\frac{16 \alpha ^5}{63}+\frac{80 \alpha ^7}{231}+\frac{160 \alpha ^9}{429}+\frac{160 \alpha ^{11}}{429}+\cdots
\right) 
\nonumber\\
&= G_{rr'}|_{\ell=4}.
\end{align}

\begin{align}
\langle 0| F_{\hat t \hat r} (t,  r,  \theta, \phi)   F_{\hat t' \hat r'} ( t, r',  \theta', \phi ') |0\rangle _{|\ell=5}&=   
  \frac{Y_{5  m}(\theta, \phi)Y_{5 m}^{*}(\theta', \phi ')}{\pi ( rr')^2} \left( 
 \frac{40 \alpha ^6}{231}+\frac{40 \alpha ^8}{143}+\frac{48 \alpha ^{10}}{143}+\frac{80 \alpha ^{12}}{221 }\cdots
 \right)
 \nonumber\\
&= G_{rr'}|_{\ell=5}.
\end{align}

\bibliographystyle{JHEP}
\bibliography{references}

\end{document}